\documentclass[12pt]{article}
\usepackage{amssymb}
\usepackage{amsmath}

\usepackage{graphicx}
 
\ifx\pdfoutput\undefined
     \relax
\else
     \usepackage{epstopdf}
\fi

\makeatletter
\def\numberbysection{\@addtoreset{equation}{section}
 	\def\theequation{\thesection.\arabic{equation}}}
\makeatother

\numberbysection

\newcommand{\be}{\begin{eqnarray}}
\newcommand{\ee}{\end{eqnarray}}
\newcommand{\non}{\nonumber}

\newcommand{\rb}[1]{\raisebox{1.5ex}[0pt]{#1}}

\newcommand{\af}{\ensuremath{\mathsf{a}}}
\newcommand{\Af}{\ensuremath{\mathsf{A}}}
\newcommand{\bff}{\ensuremath{\mathsf{b}}}
\newcommand{\Bf}{\ensuremath{\mathsf{B}}}
\newcommand{\Gf}{\ensuremath{\mathsf{G}}}
\newcommand{\Pf}{\ensuremath{\mathsf{P}}}
\newcommand{\yf}{\ensuremath{\mathsf{y}}}
\newcommand{\Yf}{\ensuremath{\mathsf{Y}}}

\def\ga{\gamma}

\def\ep{\epsilon}

\def\la{\lambda}
\def\th{\theta}

\def\La{\Lambda}

\newcommand{\beq}{\begin{equation}}
\newcommand{\eeq}{\end{equation}}
\newcommand{\bea}{\begin{eqnarray*}}
\newcommand{\eea}{\end{eqnarray*}}
\newcommand{\beqa}{\begin{eqnarray}}
\newcommand{\eeqa}{\end{eqnarray}}

\begin{document}

\begin{titlepage}
\strut\hfill UMTG--251
\vspace{.5in}
\begin{center}

\LARGE Finite size effects in the spin-1 XXZ\\
       and supersymmetric sine-Gordon models\\
\LARGE with Dirichlet boundary conditions \\[1.0in]
\large Changrim Ahn \footnote{
       Department of Physics, Ewha Womans University, 
       Seoul 120-750, South Korea (permanent address)}${}^{, 2}$,
       Rafael I. Nepomechie \footnote{
       Physics Department, P.O. Box 248046, University of Miami,
       Coral Gables, FL 33124 USA}
   and Junji Suzuki \footnote{
       Department of Physics, Faculty of Science, Shizuoka University,
       Ohya 836, Shizuoka, Japan}\\

\end{center}

\vspace{.5in}

\begin{abstract}
Starting from the Bethe Ansatz solution of the open integrable spin-1
XXZ quantum spin chain with diagonal boundary terms, we derive a set
of nonlinear integral equations (NLIEs), which we propose to describe
the boundary supersymmetric sine-Gordon model BSSG${}^{+}$ with
Dirichlet boundary conditions on a finite interval.  We compute the
corresponding boundary $S$ matrix, and find that it coincides with the
one proposed by Bajnok, Palla and Tak\'acs for the Dirichlet
BSSG${}^{+}$ model.  We derive a relation between the (UV) parameters
in the boundary conditions and the (IR) parameters in the boundary $S$
matrix.  By computing the boundary vacuum energy, we determine a previously
unknown parameter in the scattering theory.  We solve the NLIEs
numerically for intermediate values of the interval length, and find
agreement with our analytical result for the effective central charge
in the UV limit and with boundary conformal perturbation theory.
\end{abstract}

\end{titlepage}

\setcounter{footnote}{0}

\section{Introduction}\label{sec:intro}

Much can be learned from studying quantum field theories in finite
volume.  This is particularly true for $1+1$-dimensional integrable
QFTs, for which there are effective descriptions in both the infrared
(infinite size) and ultraviolet (zero size) limits, namely, massive
factorized scattering theory \cite{ZZ, RSOS, GZ} and conformal field theory
(CFT) \cite{BPZ, CFT}, respectively.  Moreover, at least for some
examples, there also exist effective descriptions in terms of certain
nonlinear integral equations (NLIEs) for {\it general} system size.

A case in point is the sine-Gordon (SG) model on a circle (periodic
boundary conditions) \cite{KP}-\cite{Ra}, and on an interval with
either Dirichlet \cite{LMSS, ABR} or general integrable \cite{AN} boundary
conditions.  NLIEs have been obtained in these papers from Bethe
Ansatz solutions of corresponding critical spin-$\frac{1}{2}$ XXZ chains
\cite{periodic}-\cite{nondiagonal}, with a mass scale introduced by
means of an alternating inhomogeneity parameter ($\pm \Lambda$).
Among the quantities that have been computed from these NLIEs are bulk
and boundary $S$ matrices (IR limit), bulk and boundary energies, and
central charge and conformal dimensions (UV limit).  Moreover, Casimir
energies in the near UV region obtained by numerically solving the NLIEs
agree well with those obtained using the truncated conformal space
approach (TCSA) \cite{YZ, DPTW} and conformal perturbation theory.

\begin{figure}[htb]
	\centering
	\includegraphics[width=0.30\textwidth]{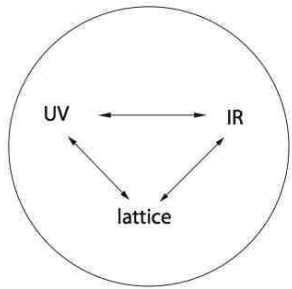}
	\caption[xxx]{\parbox[t]{0.5\textwidth}{
Three sets of parameters and their relations}
	}
	\label{fig:parameters}
\end{figure}

In the NLIE approach one generally deals with three sets of
parameters: the UV parameters appearing in the action, the IR
parameters appearing in the $S$ matrices, and the lattice parameters
in terms of which the NLIE is initially formulated.  (See Figure
\ref{fig:parameters}.)  In principle, by matching the UV and IR limits
of the NLIE with corresponding known results, one can deduce the
``lattice $\leftrightarrow$ UV'' and ``lattice $\leftrightarrow$ IR''
relations, respectively.  If the ``UV $\leftrightarrow$ IR'' relation
is also known, then the consistency of these three sets of relations
can be checked.  For the sine-Gordon model, for which the ``UV
$\leftrightarrow$ IR'' relation for the boundary parameters has been
determined \cite{uvir}, this consistency has been established for both
the bulk and boundary parameters.

Much less is known about the supersymmetric sine-Gordon (SSG) model
\cite{ssg1}-\cite{ABL}, \footnote{We take 
$\gamma^{0}=\left( \begin{array}{cc}
0 & i \\
-i & 0
\end{array} \right)\,, \quad \gamma^{1}=\left( \begin{array}{cc}
0 & i \\
i & 0
\end{array} \right)\,, \quad \Psi= \left( \begin{array}{c}
 \psi \\
 \bar \psi
\end{array} \right)$, with $\psi$ and $\bar\psi$ real.}
\be 
{\cal L}={1\over 2}\partial_{\mu}\varphi\, \partial^{\mu}\varphi  
+ {1\over 2}i \bar \Psi \gamma^{\mu} \partial_{\mu} \Psi 
- {m_{0}\over 2} \cos(\beta \varphi) \bar \Psi \Psi
+ {m_{0}^{2}\over 2 \beta^{2}} \cos^{2} (\beta \varphi) 
\,,  \label{ssg} 
\ee 
where $\varphi$ is a real scalar field, $\Psi$ is a Majorana Fermion
field, $m_{0}$ is a mass parameter, and $\beta \in \left(0 \,,
4\pi\right)$ is a dimensionless coupling constant.  Indeed, for the
case of periodic boundary conditions, an NLIE was proposed only
recently \cite{Du}, and derived in \cite{Su1, Su2} from the Bethe
Ansatz solution of the integrable spin-1 XXZ chain \cite{ZF, spin1BA}.
While the ground state of the critical spin-$\frac{1}{2}$ chain is
described by a sea of real Bethe roots, the ground state of the
critical spin-1 chain is described by a sea of approximate
``two-strings'', i.e., certain complex conjugate pairs of Bethe roots.
As a result, the familiar method \cite{DdV} of deriving the NLIE,
based on the Bethe Ansatz equations and the corresponding counting
function, does not seem to work for the spin-1 case.  Nevertheless, an
NLIE can be derived \cite{Su1, Su2} from the model's $T-Q$ equations,
in a manner similar to the original approach \cite{KP}.  Very
recently, the periodic spin-1 XXZ/SSG NLIE for excited states was
shown \cite{HRS} to have the correct UV and IR limits.  In particular,
the bulk soliton $S$ matrix \cite{ABL} was obtained from the IR limit
of the NLIE.

Our goal has been to find an NLIE for the boundary supersymmetric
sine-Gordon model on an interval with general integrable boundary
conditions \cite{IOZ}-\cite{BPT}.  This field theory is of interest as
a toy model with boundary that is both integrable and supersymmetric;
and it may also have applications to superstring theory \cite{string}.
Since a Bethe Ansatz solution of the corresponding open spin-1 XXZ
chain with general integrable boundary terms \cite{IOZ2} is still not
known, we focus here on the special case of diagonal boundary terms,
whose solution is already available \cite{MNR}.  The homogeneous chain
has the local Hamiltonian
\be
H =  \sum_{n=1}^{N-1} H_{n,n+1}  + \ b.t.  \,, \label{hamiltonian}
\ee
where the bulk terms $H_{n,n+1}$ are those of the Fateev-Zamolodchikov
\cite{ZF} spin chain,
\be 
H_{n,n+1} &=&  \sigma_{n} - (\sigma_{n})^{2}
- 2 \sin^2 \ga \left[ \sigma_{n}^{z} + (S^z_n)^2
+ (S^z_{n+1})^2 - (\sigma_{n}^{z})^2 \right] \non \\
&+& 4 \sin^2 (\frac{\ga}{2})  \left( \sigma_{n}^{\bot} \sigma_{n}^{z}
+ \sigma_{n}^{z} \sigma_{n}^{\bot} \right) \,, \label{bulkhamiltonian}
\ee 
where
\be
\sigma_{n} = \vec S_n \cdot \vec S_{n+1} \,, \quad
\sigma_{n}^{\bot} = S^x_n S^x_{n+1} + S^y_n S^y_{n+1}  \,, \quad
\sigma_{n}^{z} = S^z_n S^z_{n+1} \,, 
\ee 
and $\vec S_n$ are spin-1 generators of $SU(2)$; and the diagonal boundary 
terms are give by
\be 
b.t. &=& \frac{1}{2}\sin(2\ga) \Big\{ 
-\left[ \cot \eta_- + \cot ( \eta_- - \ga) \right]S^z_1 + 
\left[ \cot \eta_- - \cot ( \eta_- - \ga) \right](S^z_1)^2 \non \\
&- & \left[ \cot \eta_+ + \cot ( \eta_+ - \ga) \right]S^z_N
   + \left[ \cot \eta_+ - \cot ( \eta_+ - \ga) \right](S^z_N)^2 
\Big\}  \,.
\label{diagbt}
\ee 
The bulk and boundary parameters are $\ga$ and $\eta_{\pm}$, 
respectively.

We propose that, in analogy with the spin-$\frac{1}{2}$ XXZ/SG model, 
this open spin chain corresponds to the boundary SSG model on an interval
$\left[ x_{-} \,, x_{+}\right]$ with Dirichlet boundary conditions,
\be
\varphi(x_{-},t)=\varphi_{-}\,, \quad \varphi(x_{+},t)=\varphi_{+}\,, 
\quad \psi(x_{-},t) - \bar\psi(x_{-},t) = 0  \,, 
\quad \psi(x_{+},t) - \bar\psi(x_{+},t) = 0  \,, 
\label{Dirichlet}
\ee
where $\psi$ and $\bar \psi$ are the spinor components of the Majorana 
field $\Psi$. These boundary conditions follow from the boundary action
\cite{Ne} in the limit that the boundary mass parameters tend to
infinity. That the diagonal boundary terms (\ref{diagbt}) of the 
spin-1 XXZ 
chain correspond to the Dirichlet boundary conditions (\ref{Dirichlet}) 
of the SSG model is consistent 
with the fact that $S^{z}$ and topological charge are conserved in the 
two models, respectively, and also that both models are integrable.

There are actually two known sets of integrable supersymmetric
boundary conditions for the boundary SSG model \cite{Ne}.  Following
\cite{BPT}, we shall refer to the set (\ref{Dirichlet}) as Dirichlet
BSSG${}^{+}$, and to the set with $\psi + \bar \psi = 0$ at both ends
as Dirichlet BSSG${}^{-}$.

We derive an NLIE for the Dirichlet BSSG${}^{+}$ model, circumventing
(as in \cite{Su2}) the difficulties posed by the ground-state sea of
two-strings by identifying suitable auxiliary functions from the
model's $T-Q$ equations \cite{MNR}, and exploiting their analytic
properties.  By analyzing the IR limit of this NLIE, we compute the
soliton boundary $S$ matrix, which coincides with the one proposed for
the Dirichlet BSSG${}^{+}$ model by Bajnok {\it et al.} \cite{BPT}.
We propose the ``UV $\leftrightarrow$ IR'' relation for the boundary
parameters, for the special case of Dirichlet boundary conditions, on
the basis of our NLIE and its UV and IR limits.  By computing the
boundary vacuum energy, we determine a previously unknown parameter in the
scattering theory \cite{BPT}. We solve the NLIEs
numerically for intermediate values of the volume, and find agreement
with our analytical result for the effective central charge in the UV
limit and with boundary conformal perturbation theory, and confirm 
the UV-IR relation.

The outline of this paper is as follows.  
In Section \ref{sec:spinhalf} we rederive the NLIE for the
spin-$\frac{1}{2}$ XXZ/sine-Gordon model with Dirichlet boundary
conditions \cite{LMSS}.  However, we use the method which we apply to
the spin-1 case (which differs from the
approach used in \cite{LMSS}), and therefore, this serves as a
valuable warm-up exercise for the latter problem.
In Section \ref{sec:spin1} we turn to our main interest,
the spin-1 XXZ/supersymmetric sine-Gordon model
with Dirichlet boundary conditions. We
sketch the derivation of the NLIE, relegating some of the details
to Appendix \ref{sec:boundterms}.  In particular, we determine
the boundary terms $\Pf_{bdry}(\theta)$ and $\Pf_{y}(\theta)$
which encode boundary effects. Section \ref{sec:irlimit} is
devoted to an analysis of the IR limit of this NLIE. In the course of 
computing the corresponding boundary $S$ matrix, we determine the ``lattice
$\leftrightarrow$ IR'' relation for the boundary parameters.
In Section \ref{sec:uvlimit}, we analyze the UV
limit of our NLIE, and compare with the expected CFT result.
In this way, we obtain a boundary ``UV $\leftrightarrow$ lattice'' 
relation, and therefore finally the ``UV $\leftrightarrow$ IR''
relation for the boundary parameters.  In Section
\ref{sec:intermediate} we compute the effective central charge of
the Dirichlet SSG model using
first-order boundary conformal perturbation theory, and compare with
numerical NLIE results. We conclude in Section \ref{sec:conclude} with a
discussion of our results and with some comments on various open problems. 
Some important technical details are explained in the appendices.
 
\section{Spin-$\frac{1}{2}$ XXZ/SG with Dirichlet boundary conditions}\label{sec:spinhalf}

We rederive here the NLIE for the spin-$\frac{1}{2}$ XXZ/sine-Gordon model
with Dirichlet boundary conditions \cite{LMSS}.  However, in contrast
to the familiar approach \cite{DdV} used in \cite{LMSS}, we do not
introduce the counting function.  Instead, we identify suitable
auxiliary functions from the model's $T-Q$ equation, and exploit 
their analyticity properties.  We shall use the same method to treat
the spin-1 XXZ/supersymmetric sine-Gordon model in Section \ref{sec:spin1}.

\subsection{$T-Q$ equation}

The transfer-matrix eigenvalues $T(x)$ of the inhomogeneous open
spin-$\frac{1}{2}$ XXZ chain with Dirichlet boundary conditions
satisfy the $T-Q$ equation \cite{Sk}
\be 
T(x)=T^{(+)}(x)+T^{(-)}(x) \,, \qquad 
T^{(\pm)}(x)\equiv \sinh(2x \pm i \ga)\, B^{(\pm)}(x)\, \phi(x\pm\frac{i\ga}{2})
\frac{Q(x\mp i\ga)}{Q(x)} \,,
\label{TQspin12}
\ee
where 
\be
\phi(x)&=& \sinh^N(x-\La) \sinh^N(x+\La)  \,, \non \\
B^{(\pm)}(x)&=&\sinh (x\pm \frac{i\ga H_{+}}{2})
	     \sinh (x\pm \frac{i\ga H_{-}}{2}) \,, \non \\
Q(x)&=&\prod_{k=1}^{M}\sinh(x-v_k) \sinh(x+v_k) \,.
\ee
We denote the bulk parameter by $\ga$ ($0 < \ga < \pi$),
and the boundary parameters by $H_{\pm}$.
Moreover, $\La$ is the inhomogeneity parameter which provides a
mass scale; $N$ is the number of spins; and the $M$ zeros $v_k$ of
$Q(x)$ are the Bethe roots. Note that $T(-x) = -T(x)$.

For the homogeneous case $\La=0$, a local Hamiltonian is obtained 
from the first derivative of the transfer matrix $T(x)$ \cite{Sk}. However, 
for the inhomogeneous case $\La \ne 0$ which we consider here, the definition of 
energy is less clear. We shall follow the prescription of Reshetikhin 
and Saleur \cite{RS}, which implies 
\beq
E=-\frac{g}{a}\left\{ \frac{d}{dx}\ln T(x)\Bigg\vert_{x=\Lambda+\frac{i\gamma}{2}}
-\frac{d}{dx}\ln T(x)\Bigg\vert_{x=\Lambda-\frac{i\gamma}{2}}\right\} 
\,, \label{energydef}
\eeq
where $a$ is the lattice spacing, and $g$ is given by 
\be
g=-\frac{i\ga}{4\pi} \,. \label{energynormalization}
\ee 
One can verify that this definition has the correct $\La \rightarrow 0$ limit.

\subsection{Derivation of NLIE}

We define the auxiliary functions $a(x)$ and $\bar a(x)$ by
\beq
a(x)={\sinh(2x+i\ga)\, B^{(+)}(x)\, \phi(x+\frac{i\ga}{2})\, Q(x-i\ga)\over{
      \sinh(2x-i\ga)\, B^{(-)}(x)\, \phi(x-\frac{i\ga}{2})\, Q(x+i\ga)}} \,, \qquad
      {\bar a}(x)=a(-x)={1\over a(x)}
\,.
\label{defa}
\eeq
The transfer-matrix eigenvalues can then be written as
\be
T(x) &=& \sinh(2x-i\ga)\, 
B^{(-)}(x)\, \phi(x-\frac{i\ga}{2}){Q(x+i\ga)\over{Q(x)}}A(x) \non \\
&=&\sinh(2x+i\ga)\, 
B^{(+)}(x)\, \phi(x+\frac{i\ga}{2}){Q(x-i\ga)\over{Q(x)}}{\bar 
A}(x) \,,
\label{TQhalf}
\ee
where
\beq
A(x)=1+a(x) \,,\quad {\bar A}(x)=1+{\bar a}(x)  \,.
\eeq
The Bethe Ansatz equations are given by
\be
A(v_{k}) = 0 \,, \qquad k = 1\,, \ldots \,, M \,.
\ee 

We consider the ground state. For simplicity, we restrict the 
boundary parameters $H_{\pm}$ to the interval 
\be
0 < H_{\pm} <  \frac{\pi}{\ga} \,.
\label{bpdomainspin120}
\ee 
We argue in Appendix \ref{sec:domainspinhalf}
that the boundary parameters should be further restricted to the range 
\be 
\frac{\pi}{\ga}-2 < H_{+} + H_{-} < \frac{3\pi}{\ga}-2 
\label{bpdomainspin12}
\ee 
in order for the ground state to have $M=N/2$ real Bethe roots 
$v_{k}$ and no holes, except for one hole at the origin.
That is, $T(x)$ does not have zeros near the real
axis except for a simple zero at the origin. To remove this root, we define
\beq
\check{T}(x)={T(x)\over \mu(x)} \,, 
\eeq
where $\mu(x)$ is any function whose only real root is a simple zero at the origin,
in particular $\mu(0)=0\,, \ \mu'(0)\ne 0$, so that $\check{T}(x)$ is analytic
and nonzero (ANZ) when $x$ is near the real axis.  (We use a prime $(\
'$) to denote differentiation with respect to $x$.) It is convenient 
to introduce the compact notation
\be
\check{T}(x) = t_{-}(x)\, {Q(x+i\ga)\over{Q(x)}}A(x) = 
t_{+}(x)\, {Q(x-i\ga)\over{Q(x)}} \bar A(x) \,,
\ee
where
\be
t_{\pm}(x)={\sinh(2x \pm i\ga )\over \mu(x)}B^{(\pm)}(x)\, 
\phi(x \pm \frac{i\ga}{2}) \,.
\ee 

\begin{figure}[htb]
    \centering
    \includegraphics[height=5cm]{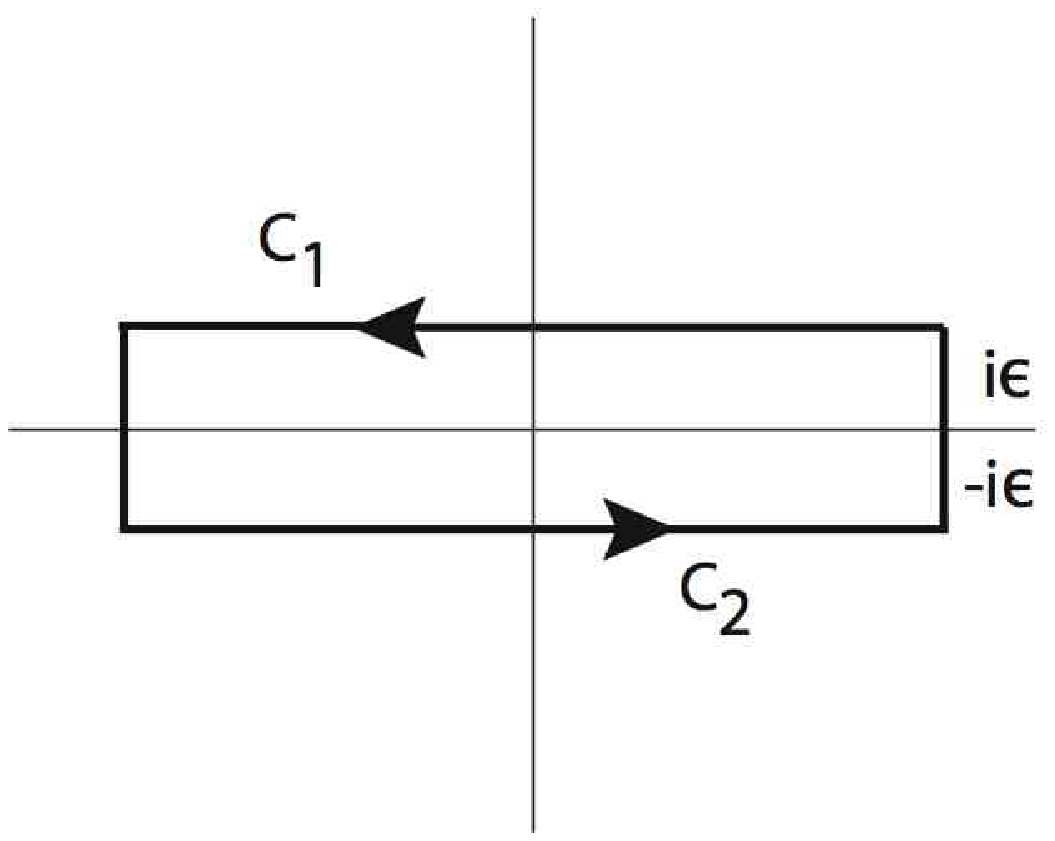}
    \caption[xxx]{\parbox[t]{0.25\textwidth}{
Integration contour}
	}
	\label{fig:contour}
\end{figure}

Since $\ln \check{T}(x)$ is analytic near the real axis, Cauchy's theorem gives
\footnote{Since $\check{T}(x)$ grows exponentially with $x$ for large 
$x$ (as follows from (\ref{TQspin12})), $[\ln \check{T}(x)]''\rightarrow 0$ for 
$x\rightarrow \infty$.}
\beq
0=\oint_C dx\ [\ln\check{T}(x)]'' e^{ikx} \,,
\eeq
where we choose the contour $C$ as in Figure \ref{fig:contour}.

Dividing the contour $C=C_1+C_2$, where $C_{1}$ and $C_{2}$ have 
imaginary parts $\pm i\epsilon$ with $\epsilon$ small and positive 
respectively, this integral can be written as
\bea
0&=&\int_{C_1} dx\ \left[ \ln t_{+}(x) \right]'' e^{ikx}+
\int_{C_1} dx\ \left\{ \ln \left[{Q(x-i\ga)\over Q(x)}\right] \right\}'' e^{ikx}
+\int_{C_1} dx\ \left[ \ln {\bar A}(x) \right]'' e^{ikx} \\
&+& \int_{C_2} dx\ \left[ \ln t_{-}(x) \right]'' e^{ikx} 
+ \int_{C_2} dx\ \left\{ \ln\left[{Q(x+i\ga)\over Q(x)}\right] 
\right\}'' e^{ikx}
+\int_{C_2} dx\ \left[ \ln A(x)\right]'' e^{ikx} \,.
\eea
In terms of Fourier transforms defined along $C_{2}$ and $C_{1}$ by 
\be
\widehat{Lf''}(k)=\int_{C_2} dx\ [\ln f(x)]'' e^{ikx} \,, \qquad
\widehat{{\cal L}f''}(k)=\int_{C_1} dx\ [\ln f(x)]'' e^{ikx} \,,
\label{fouriertransfdef}
\ee
respectively, we can rewrite
\be
\int_{C_1}dx\ \left\{ \ln \left[{Q(x-i\ga)\over Q(x)}\right]
\right\}''e^{ikx}&=&
-\widehat{LQ''}(k)\left(e^{-\ga k}-e^{-\pi k}\right) \,, \non \\
\int_{C_2}dx\ \left\{ \ln \left[{Q(x+i\ga)\over Q(x)}\right] 
\right\}'' e^{ikx}&=&
\widehat{LQ''}(k)\left[e^{(\ga-\pi)k}-1\right] \,,
\ee
where we have used the periodicity
\beq
Q(x)=Q(x-i\pi),\quad x\in C_1,\qquad{\rm and}\qquad
Q(x+i\ga)=Q(x+i\ga-i\pi),\quad x\in C_2 \,.
\label{Qperiodicity}
\eeq
Defining
\beqa
C(k) \equiv \int_{C_1}dx \left[ \ln t_{+}(x) \right]'' e^{ikx}
+\int_{C_2}dx \left[ \ln t_{-}(x) \right]'' e^{ikx} \,,
\label{Ck}
\eeqa
we obtain
\beq
\widehat{LQ''}(k)={e^{ \frac{\pi k}{2}}\over 4 \cosh( \frac{\ga k}{2})
\sinh\left((\pi-\ga)\frac{k}{2}\right)}
\left[
\widehat{LA''}(k)+\widehat{{\cal L}{\bar A}''}(k)+C(k)\right] \,.
\label{logQ}
\eeq
This is the main consequence of analyticity.

It follows from the definition of $a(x)$ (\ref{defa}) that
\beq
\widehat{La''}(k)=\int_{C_2}dx\ \left\{
\ln\left[{Q(x-i\ga)\over Q(x+i\ga)}\right]\right\}'' e^{ikx} + D(k) 
\,,
\eeq
where $D(k)$ is defined by
\beq
D(k)=\int_{C_2}dx\
\left\{ \ln\left[{\sinh(2x+i\ga)\, B^{(+)}(x)\, \phi(x+\frac{i\ga}{2})\over
\sinh(2x-i\ga)\, B^{(-)}(x)\, \phi(x-\frac{i\ga}{2})}\right] 
\right\}'' e^{ikx} \,.
\label{dkdef}
\eeq
Using again the periodicity of $Q(x)$, we obtain
\beq
\widehat{La''}(k)=\widehat{LQ''}(k)\left[e^{-\ga 
k}-e^{(\ga-\pi)k}\right]+D(k) \,.
\label{loga}
\eeq
Inserting (\ref{logQ}) into (\ref{loga}) yields
\beq
\widehat{La''}(k)= {\widehat G}(k)\left[\widehat{LA''}(k)+\widehat{{\cal L}{\bar 
A}''}(k)\right]+ C_{T}(k) \,, \label{nlie}
\eeq
where
\be
{\widehat G}(k) &=&{\sinh\left((\pi-2\ga)\frac{k}{2}\right)
\over 2\cosh (\frac{\ga k}{2})\sinh\left((\pi-\ga)\frac{k}{2}\right)} 
\,, \label{Gk} \\
C_{T}(k) &=&{\widehat G}(k)\, C(k)+D(k) \,. \label{CTdef}
\ee
The evaluation of $C_{T}(k)$ is tedious but
straightforward. To this end, we make use of the identities 
\beq
\int_{C_2} {dx\over 2\pi}  \left[ \ln \sinh(x-i \alpha)\right]'' e^{ikx}= 
e^{-k(\alpha - n \pi)}  \psi(k) \,, \qquad 
\psi(k) \equiv {k\over 1-e^{-\pi k}} \,, 
\label{psi}
\eeq
where $n$ is an integer such that $0 < \Re e(\alpha - n \pi) < \pi$, 
and
\beq
\int_{C_2} {dx\over 2\pi} \left[ \ln \sinh(2x)\right]'' e^{ikx}
={k\over 1-e^{-\frac{\pi k}{2}}} \equiv \psi_2(k) \,.
\label{psi2}
\eeq
The result is (\ref{CTfinal})
\be
C_{T}(k) &=& -2\pi k \Bigg\{ {N\cos(\La k)\over \cosh({\ga k\over 2})} 
+{\sinh\left((\ga H_{+}-\pi)\frac{k}{2}\right)
   +\sinh\left((\ga H_{-}-\pi)\frac{k}{2}\right)
\over 2\cosh({\ga k\over 2}) \sinh\left((\ga-\pi)\frac{k}{2}\right)} 
\non \\
& & +{\cosh({\ga k\over 4}) \sinh\left((2\ga-\pi)\frac{k}{4}\right)\over
  \cosh({\ga k\over 2}) \sinh\left((\ga-\pi)\frac{k}{4}\right)} 
\Bigg\} \,. 
\ee

The result (\ref{nlie}) is the NLIE for the lattice sine-Gordon model with Dirichlet 
boundary conditions in Fourier space.  Passing to coordinate space, and integrating twice,
we obtain 
\be
\ln a(x) &=& 
\int_{-\infty}^{\infty}dx'\ G(x-x'+i\epsilon) \ln A(x'-i\epsilon) - 
\int_{-\infty}^{\infty}dx'\ G(x-x'-i\epsilon) \ln \bar 
A(x'+i\epsilon)\non \\
&-& i 2N \tan^{-1}\left({\sinh \frac{\pi x}{\ga}\over 
\cosh \frac {\pi \La}{\ga} } \right) + i\, P_{bdry}(x) +i\pi \,,
\label{spinhalflatticeNLIE}
\ee
where $G(x)$ is the Fourier transform of $\widehat G(k)$ (\ref{Gk})
\be
G(x) = {1\over 2\pi} \int_{-\infty}^{\infty}dk\ e^{-i k x}\ \widehat 
G(k) \,,
\label{FTdefG}
\ee
$P_{bdry}(x)$ is given by
\be
P_{bdry}(x) = \int_{0}^{x}dx'\, R(x') = \frac{1}{2} 
\int_{-x}^{x}dx'\, R(x')\,,
\label{spinhalfPbdry}
\ee
and $R(x)$ is the Fourier transform of $\hat R(k)$,
\be
\hat R(k) &=& -2 \pi \Bigg\{ 
{\sinh\left((\ga H_{+}-\pi)\frac{k}{2}\right)
   +\sinh\left((\ga H_{-}-\pi)\frac{k}{2}\right)
\over 2\cosh({\ga k\over 2}) \sinh\left((\ga-\pi)\frac{k}{2}\right)} 
\non \\
& & \quad +{\cosh({\ga k\over 4}) \sinh\left((2\ga-\pi)\frac{k}{4}\right)\over
  \cosh({\ga k\over 2}) \sinh\left((\ga-\pi)\frac{k}{4}\right)} \Bigg\} \,.
\ee 
The integration constant is explained in Section \ref{sec:spinhalfintegconst}.

The continuum limit consists of taking $\La \rightarrow \infty$ 
(leading to a simplification in the driving term, 
$-i2N \tan^{-1}\left({\sinh \frac{\pi x}{\ga}\over 
\cosh \frac {\pi \La}{\ga} } \right) \sim -i4N e^{-\frac {\pi \La}{\ga}} 
\sinh \frac{\pi x}{\ga}$), together with $N
\rightarrow \infty$ and lattice spacing $a \rightarrow 0$, 
such that the interval length $L\equiv x_{+}-x_{-}$ and the soliton mass $m$
are given by
\be
L = N a \,, \qquad m={2\over a} e^{-\frac{\pi \Lambda}{\ga}} \,.
\label{continuumlimit}
\ee
The driving term therefore becomes $-i 2mL \sinh \theta$, 
where the renormalized rapidity $\theta$ is defined as
\be 
\theta = \frac{\pi x}{\gamma} \,.
\label{renormrapidity}
\ee 
The resulting NLIE 
\be
\ln \af(\theta) &=& 
\int_{-\infty}^{\infty}d\theta'\ \Gf (\theta-\theta'+i\varepsilon) \ln 
\Af(\theta'-i\varepsilon) - 
\int_{-\infty}^{\infty}d\theta'\ \Gf(\theta-\theta'-i\varepsilon) \ln \bar 
\Af(\theta'+i\varepsilon)\non \\
&-& i 2mL \sinh \theta + i\, \Pf_{bdry}(\theta) +i\pi\,,
\ee
where we have defined
\be
\varepsilon= \frac{\pi \epsilon}{\ga}\,, \quad \!
\af(\theta)=a(\frac{\ga \theta}{\pi})\,, \quad \!
\Af(\theta)=A(\frac{\ga \theta}{\pi})\,, \quad  \!
\Pf_{bdry}(\theta)=P_{bdry}(\frac{\ga \theta}{\pi})\,, \quad \!
\Gf(\theta)=\frac{\ga}{\pi}G(\frac{\ga \theta}{\pi})\,, 
\label{mathfrakdefs}
\ee 
agrees with \cite{LMSS}.  

\subsection{Vacuum and Casimir energies}\label{subsec:spinhalfvacuumCasimir}

Since our approach avoids introducing the counting function and the
density of Bethe roots, the energy computation also differs from that
of the conventional approach \cite{DdV}.  The main idea is to express the
energy in terms of $\widehat{LQ}(k)$, and then make use of the
consequence (\ref{logQ}) of analyticity.

The energy is given by (\ref{energydef}).
From the $T-Q$ equation (\ref{TQspin12}) and the fact $\phi(\La)=0$, we see that
\be 
\frac{d}{dx}\ln T(x)\Bigg\vert_{x=\Lambda\pm\frac{i\gamma}{2}}=
\frac{d}{dx}\ln 
T^{(\pm)}(x)\Bigg\vert_{x=\Lambda\pm\frac{i\gamma}{2}} \,.
\ee
Hence, the energy (\ref{energydef}) can be written as
\be
E&=&-\frac{g}{a}\frac{d}{dx}\left\{ \ln T^{(+)}(x+\frac{i\gamma}{2})-
\ln T^{(-)}(x-\frac{i\gamma}{2})\right\} \bigg\vert_{x=\Lambda} \non 
\\
&=&-\frac{g}{a}\int \frac{dk}{2\pi} e^{-ik\Lambda}\left[
e^{\frac{\ga k}{2}}\widehat{LT^{(+)'}}(k)-e^{-\frac{\ga k}{2}}\widehat{LT^{(-)'}}(k)\right] \,,
\label{FTenergy}
\ee
where we used the fact (see (\ref{fouriertransfdef}))
\be
[\ln f(x)]' = \int \frac{dk}{2\pi}\ \widehat{Lf'}(k)\ e^{-ik x} \,, \qquad
x \in C_{2} \,.
\label{FTfact}
\ee 
Since
\beq
T^{(\pm)}(x\pm\frac{i\gamma}{2})=\sinh(2x \pm 2i\ga)\, 
B^{(\pm)}\left(x\pm\frac{i\gamma}{2}\right)\phi(x\pm i\ga)
{Q\left(x\mp\frac{i\gamma}{2}\right)\over{Q\left(x\pm\frac{i\gamma}{2}\right)}},
\eeq
we can compute the Fourier transforms $e^{\pm\frac{\ga 
k}{2}}\widehat{LT^{(\pm)'}}(k)$, \footnote{We assume here that $0<\ga 
< \pi/2$.}
\be
e^{-\frac{\ga k}{2}}\widehat{LT^{(-)'}}(k)&=&e^{-{\ga 
k\over{2}}}\widehat{LB^{(-)'}}(k)
+e^{-\ga k}\widehat{L\phi'}(k) 
+ e^{-\ga k} \frac{2\pi\psi_2(k)}{(-ik)}\non \\
&-&2e^{-\frac{\pi k}{2}}\sinh((\pi-\ga)\frac{k}{2})
\widehat{LQ'}(k) \,, \non \\
e^{\frac{\ga k}{2}}\widehat{LT^{(+)'}}(k)&=&e^{{\ga k\over{2}}}
\widehat{LB^{(+)'}}(k)+e^{(\ga-\pi)k}\widehat{L\phi'}(k) 
+ e^{(\ga-\frac{\pi}{2})k} \frac{2\pi\psi_2(k)}{(-ik)}\non \\
&+& 2e^{-\frac{\pi k}{2}}\sinh((\pi-\ga)\frac{k}{2})
\widehat{LQ'}(k) 
\,.
\ee

We can eliminate $\widehat{LQ'}(k)$ using (\ref{logQ}) and the fact 
$\widehat{Lf'}(k) = {1\over (-i k)}\widehat{Lf''}(k)$,
\be
\widehat{LQ'}(k) &=& {1\over (-i k)}{e^{ \frac{\pi k}{2}}\over 4 \cosh( \frac{\ga k}{2})
\sinh\left((\pi-\ga)\frac{k}{2}\right)}
\left[
\widehat{LA''}(k)+\widehat{{\cal L}{\bar A}''}(k)+C(k)\right] \non \\
&=& {e^{ \frac{\pi k}{2}}\over 4 \cosh( \frac{\ga k}{2})
\sinh\left((\pi-\ga)\frac{k}{2}\right)}
\Big\{
\widehat{LA'}(k)+\widehat{{\cal L}{\bar 
A}'}(k)+\widehat{LB^{(-)'}}(k)-\widehat{LB^{(+)'}}(k) \non \\
&+&
\left[e^{-{\ga k\over{2}}}-e^{\left({\ga\over{2}}-\pi\right)k}\right]
\widehat{L\phi'}(k) 
+ \left[e^{-{\ga k\over{2}}}-e^{\left({\ga\over{2}}-\frac{\pi}{2}\right)k}\right] 
\frac{2\pi\psi_2(k)}{(-ik)}
-2\pi i \Big\}
\,,
\ee
where we have passed to the second line using (see (\ref{CintermsofD}), (\ref{dkdef}))
\be
C(k) &=&\widehat{LB^{(-)''}}(k)-\widehat{LB^{(+)''}}(k)+
\left[e^{-{\ga k\over{2}}}-e^{\left({\ga\over{2}}-\pi\right)k}\right]
\widehat{L\phi''}(k) \non \\
&+&\left[e^{-{\ga k\over{2}}}-e^{\left({\ga\over 2}-{\pi\over 2}\right)k}\right]
2\pi\psi_2(k) -\delta(k) \,.
\ee
We obtain
\be
\lefteqn{
e^{\frac{\ga k}{2}}\widehat{LT^{(+)'}}(k)-e^{-\frac{\ga k}{2}}\widehat{LT^{(-)'}}(k) =
\frac{1}{\cosh{\ga k\over{2}}}\left[
\widehat{LA'}(k)+\widehat{{\cal L}{\bar A}'}(k)-2\pi i\right]} \non \\
& & +\tanh({\ga k\over{2}})\bigg[
2e^{-{\pi k\over{2}}}\cosh((\ga-{\pi\over{2}})k) \widehat{L\phi'}(k) 
- {2\pi\cosh((\ga-\frac{\pi}{4})k)\over i\sinh \frac{\pi k}{4}}
\non \\
& & +e^{-{\ga k\over{2}}}\widehat{LB^{(-)'}}(k)+e^{{\ga 
k\over{2}}}\widehat{LB^{(+)'}}(k)
\bigg] \,. \label{FTenbergyintegrand}
\ee
This is essentially the integrand in the expression (\ref{FTenergy}) for the energy.

Let us consider this result term by term.
The $\widehat{L\phi'}(k)$ is obtained as before (\ref{Dk})
\[
\widehat{L\phi'}(k)=2\pi N(e^{i\Lambda k}+e^{-i\Lambda 
k})\frac{\psi(k)}{(-ik)} \,.
\]
Inserting the corresponding order $N$ contribution from
(\ref{FTenbergyintegrand}) into (\ref{FTenergy}), we obtain the bulk
energy
\beq
E_{B}= \frac{Ng}{i a}\int_{-\infty}^{\infty} dk\ (1+e^{-2i\Lambda k})
\frac{\sinh{\ga k\over{2}}\cosh\left((\ga-{\pi\over{2}})k\right)}
{\cosh{\ga k\over{2}}\sinh{\pi k\over{2}}} \,.
\label{FTBenergy}
\eeq
This quantity is divergent in the continuum limit.  We adopt the
renormalization procedure \cite{DdV} of discarding divergent terms and
keeping only the (finite) terms that can be expressed in terms of the
physical mass $m$ given in (\ref{continuumlimit}).  To this end, we
discard the $\Lambda$-independent term, and evaluate the remaining
integral by closing the contour in the lower half plane and selecting
only the contribution from the pole at $k=-\frac{i\pi}{\ga}$.  We
arrive at the result \cite{DdV, AZ2}
\beq
E_{B}=-\frac{4\pi g}{i\ga}\frac{N}{a}e^{-{2\pi \Lambda\over{\ga}}}
\cot\frac{\pi^2}{2\ga}
=\frac{1}{4}Lm^2\cot\frac{\pi^2}{2\ga} \,,
\eeq
where we have used (\ref{energynormalization}) and (\ref{continuumlimit}).

We now consider the boundary energy. From (\ref{Dk}), we 
have
\be
e^{-{\ga k\over{2}}}\widehat{LB^{(-)'}}(k)+e^{{\ga 
k\over{2}}}\widehat{LB^{(+)'}}(k)
&=&\frac{2\pi}{(-ik)}\psi(k)\Big[
e^{\left({\ga (H_{+}+1)\over{2}}-\pi\right)k}
+e^{\left({\ga (H_{-}+1)\over{2}}-\pi\right)k} \non \\
& & + e^{-{\ga (H_{+}+1)\over{2}}k}+e^{-{\ga 
(H_{-}+1)\over{2}}k}\Big] \,. 
\ee 
Substituting this contribution from (\ref{FTenbergyintegrand})
into  (\ref{FTenergy}), we obtain the boundary energy
\be
E_{b}&=&\frac{g}{i a}\int_{-\infty}^{\infty} dk\ e^{-i\Lambda k} \Bigg\{
\frac{\sinh{\ga k\over{2}}}{\cosh{\ga k\over{2}}\sinh{\pi k\over 2}}
\left[\cosh\left((\ga (H_{+}+1)-\pi)\frac{k}{2}\right)
+(H_{+}\to H_{-})\right] \non \\
&+& \frac{\sinh{\ga k\over 2} \cosh((\ga-\frac{\pi}{4})k)}
{\cosh{\ga k\over 2}\sinh{\pi k\over 4}} -\frac{1}{\cosh{\ga k\over 2}}
\Bigg\} \,,
\label{FTbenergy}
\ee
which can be evaluated using the same contour integral as before,
\beq
E_{b}={m\over 2}\left[1 + \cot \frac{\pi^{2}}{4 \ga}+
 \frac{\sin\left(\frac{\pi}{2}(H_{+}-\frac{\pi}{\ga})\right)}{\sin\frac{\pi^{2}}{2\ga}}
+\frac{\sin\left(\frac{\pi}{2}(H_{-}-\frac{\pi}{\ga})\right)}{\sin\frac{\pi^{2}}{2\ga}}\right] 
\,.
\eeq
This matches with the result in \cite{LMSS}. 

The Casimir energy is given by (see (\ref{FTenergy}), (\ref{FTenbergyintegrand}))
\beq
E_C=-\frac{g}{a}\int_{-\infty}^{\infty} \frac{dk}{2\pi}\  e^{-i\Lambda k}
\frac{1}{\cosh{\ga k\over{2}}}\left[
\widehat{LA'}(k)+\widehat{{\cal L}{\bar A}'}(k)\right] \,.
\eeq
Passing to coordinate space and taking the continuum limit, we obtain
\be
E_C&=&\frac{2g}{a\ga}\Im m
\int_{-\infty}^{\infty} dx\left(\frac{1}{\cosh\frac{\pi}{\ga}
(\Lambda-x+i\ep)}\right)'\ln A(x-i\ep) \non \\
&=&\frac{m}{2\ga}\Im m
\int_{-\infty}^{\infty} dx\, e^{\frac{\pi}{\ga}(x-i\ep)}
\ln A(x-i\ep)
=\frac{m}{2\pi}\Im m\int_{-\infty}^{\infty} d\th\, e^{\th-i\varepsilon}\ln 
\Af(\th-i\varepsilon) \,,
\ee
where the renormalized rapidity $\theta$ is given by (\ref{renormrapidity}).
Using ${\bar \Af}(x)=\Af(-x)$ and $\Im m z=-\Im m{\bar z}$, the last expression
can be rewritten as
\beqa
E_C&=&\frac{m}{4\pi}\Im m\int_{-\infty}^{\infty} d\th
\left[e^{\th-i\varepsilon}\ln \Af(\th-i\varepsilon)
-e^{-\th+i\varepsilon}\ln {\bar \Af}(-\th+i\varepsilon)\right]
\nonumber\\
&=&\frac{m}{2\pi}\Im m\int_{-\infty}^{\infty} d\th\
\sinh(\th-i\varepsilon)\ln \Af(\th-i\varepsilon) \,.
\eeqa

\section{Spin-1 XXZ/SSG with Dirichlet boundary conditions}\label{sec:spin1}

We turn now to our main interest, the spin-1 XXZ/supersymmetric sine-Gordon model
with Dirichlet boundary conditions.

\subsection{$T-Q$ equations}\label{subsec:TQspin1}

For the spin-1 chain, there are two relevant commuting transfer matrices: 
$T_1(x)$ with a spin-$\frac{1}{2}$ (two-dimensional) auxiliary space, 
and $T_2(x)$ with a spin-1 (three-dimensional) auxiliary space. The 
corresponding eigenvalues (which we denote by the same notation) 
obey $T-Q$ equations found in \cite{MNR}: $T_1(x)$ can be written 
as \footnote{In \cite{MNR}, the transfer matrices were defined  
with some multiplicative factors which we omit here.}
\beqa
T_1(x)&=&l_1(x)+l_2(x) \,, \label{T1}\\
l_1(x)&=&\sinh(2x+i\ga) B^{(-)}(x) \phi(x+i\ga) {Q(x-i\ga)\over Q(x)} 
\,, \non \\
l_2(x)&=&\sinh(2x-i\ga) B^{(+)}(x) \phi(x-i\ga) {Q(x+i\ga)\over Q(x)} 
\,, \non 
\eeqa
and $T_2(x)$ can be written as
\beqa
T_2(x)&=&\la_1(x)+\la_2(x)+\la_3(x) \,, \label{T2}\\
\la_1(x)&=&\sinh(2x-2i\ga) B^{(+)}(x-{i\ga\over{2}}) B^{(+)}(x+{i\ga\over{2}})
\phi(x-{3i\ga\over{2}})\phi(x-{i\ga\over{2}})
{Q(x+{3i\ga\over{2}})\over{Q(x-{i\ga\over{2}})}}\,, \non \\
\la_2(x)&=&\sinh(2x) B^{(-)}(x-{i\ga\over{2}}) B^{(+)}(x+{i\ga\over{2}})
\phi(x-{i\ga\over{2}})\phi(x+{i\ga\over{2}})
{Q(x+{3i\ga\over{2}})\over{Q(x-{i\ga\over{2}})}}
{Q(x-{3i\ga\over{2}})\over{Q(x+{i\ga\over{2}})}}\,, \non\\
\la_3(x)&=&\sinh(2x+2i\ga) B^{(-)}(x-{i\ga\over{2}}) B^{(-)}(x+{i\ga\over{2}})
\phi(x+{3i\ga\over{2}})\phi(x+{i\ga\over{2}})
{Q(x-{3i\ga\over{2}})\over{Q(x+{i\ga\over{2}})}} \,, \non
\eeqa
where 
\be
\phi(x)&=&\sinh^N(x-\La) \sinh^N(x+\La)\,, \non \\
B^{(\pm)}(x)&=&\sinh(x\pm i\eta_{+}) \sinh(x\pm i\eta_{-}) 
\,,\label{phiBpm} \\
Q(x)&=&\prod_{k=1}^{M}\sinh(x-v_k) \sinh(x+v_k) \,. \non 
\ee
We denote the bulk and boundary parameters by $\ga$ and $\eta_{\pm}$,
respectively; $\La$ is the inhomogeneity parameter which provides a
mass scale; $N$ is the number of spins; and the $M$ zeros $v_k$ of
$Q(x)$ are the Bethe roots. As we shall see below (\ref{bulkUVlat}), it 
suffices to restrict $\ga$ to the domain $0 <\ga < \frac{\pi}{2}$.
The domains $(0 \,, \frac{\pi}{3})$ and $(\frac{\pi}{3} \,, 
\frac{\pi}{2})$ correspond to ``repulsive'' and ``attractive'' 
regimes of the SSG model, respectively.
Note that $T_{2}(-x)=-T_{2}(x)$.

The fusion relation
\beq
T_1(x-{i\ga\over{2}})\, T_1(x+{i\ga\over{2}})=f(x)+T_0(x)\, T_2(x) \,,
\label{T1T1}
\eeq
where
\be
T_0(x) &=& \sinh(2x) \,, \non  \\
f(x) &=& l_2(x-{i\ga\over{2}})\, l_1(x+{i\ga\over{2}}) \label{deff} \\
&=&\sinh(2x-2i\ga) \sinh(2x+2i\ga) B^{(+)}(x-{i\ga\over{2}}) B^{(-)}(x+{i\ga\over{2}})
\phi(x-{3i\ga\over{2}})\phi(x+{3i\ga\over{2}})   \,, \non 
\ee
can be readily verified using the identities
\beqa
\la_1(x)&=&{1\over{T_0(x)}}l_2(x-{i\ga\over{2}})\, 
l_2(x+{i\ga\over{2}}) \,, \non \\
\la_2(x)&=&{1\over{T_0(x)}}l_1(x-{i\ga\over{2}})\, 
l_2(x+{i\ga\over{2}})\,,  \label{fusionidentities} \\
\la_3(x)&=&{1\over{T_0(x)}}l_1(x-{i\ga\over{2}})\, 
l_1(x+{i\ga\over{2}}) \,. \non
\eeqa

For the homogeneous case $\La=0$, the local Hamiltonian
(\ref{hamiltonian}) is obtained from the first derivative of the
transfer matrix $T_2(x)$ \cite{MNR}.
For the inhomogeneous case $\La \ne 0$, as in the spin-1/2 case 
(\ref{energydef}), we define the energy by
\beq
E=-\frac{g}{a}\left\{ \frac{d}{dx}\ln T_2(x)\Bigg\vert_{x=\Lambda+\frac{i\gamma}{2}}
-\frac{d}{dx}\ln T_2(x)\Bigg\vert_{x=\Lambda-\frac{i\gamma}{2}}\right\} 
\,, \label{energydef2}
\eeq
where $a$ is the lattice spacing, and $g$ is given by (\ref{energynormalization}). 

\begin{figure}[htb]
\centerline{
\includegraphics[height=6cm]{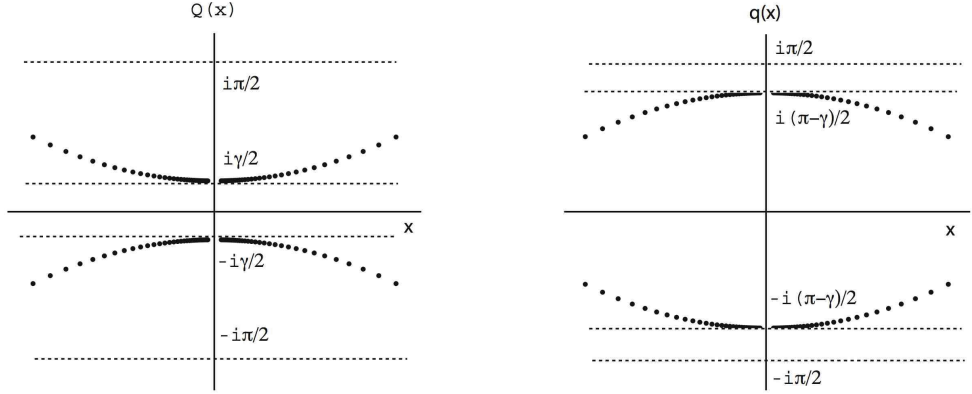}}
\caption{\small Schematic depiction of zeros of $Q(x)$ and $q(x)$, respectively}
\label{fig:zeros}
\end{figure}

We consider the ground state. For simplicity, we restrict the 
boundary parameters $\eta_{\pm}$ to the interval 
\be 
\frac{\pi}{2} < \eta_{\pm} < \pi \,.
\label{bpdomainspin10}
\ee 
We argue in Appendix \ref{sec:domainspin1}
that the boundary parameters should be further restricted to the range
\be
\frac{2\pi}{3} + \ga < \eta_{+}+\eta_{-} < \frac{4\pi}{3} + \ga 
\,,
\label{bpdomainspin1}
\ee
in order for the ground state to have $M=N$ Bethe roots $v_{k}$ 
which form approximate two-strings; that is, pairs $x_{k} \pm i 
y_{k}$ with real centers $x_{k}$ and imaginary parts $y_{k}$ 
satisfying $0< y_{k} - \frac{\ga}{2} \ll 1$,
as shown on the left-hand side of Figure
\ref{fig:zeros}. 

Since $Q(x)$ can have zeros near the real axis (namely, when $\ga$ is
close to zero), it is convenient to work instead with the shifted
quantity $q(x)$ defined by
\beq
q(x)=Q\left(x+{i\pi\over{2}}\right) \,,
\label{newqdef}
\eeq
which is ANZ near the real axis.
Also, one can check numerically that $T_1(x)$ and
$T_2(x)$ do not have zeros near the real
axis except for a simple zero at the origin.
To remove this root, we define 
\beq
\check{T}_1(x)={T_1(x)\over \mu(x)} \,, \qquad
\check{T}_2(x)={T_2(x)\over \mu(x)} \,, 
\label{checkTdef}
\eeq
where again $\mu(x)$ is any function whose only real root is a simple
zero at the origin, in particular $\mu(0)=0\,, \ \mu'(0)\ne 0$, so
that $\check T_1(x)$ and $\check T_2(x)$ are ANZ when $x$ is near the
real axis.

\subsection{Auxiliary functions}\label{subsec:auxfuncspin1}

It is convenient to define the auxiliary functions \cite{Su2}
\be
b(x)={\la_1(x)+\la_2(x)\over{\la_3(x)}}\,, \qquad
{\bar b}(x)={\la_3(x)+\la_2(x)\over{\la_1(x)}} = b(-x) \,.
\label{defb}
\ee
Since $\bar b(x)$ is the complex conjugate of $b(x)$ for real $x$, we
shall generally refrain from writing equations for $\bar b(x)$, as they can be
readily obtained by conjugation of corresponding equations for $b(x)$.
With the help of (\ref{fusionidentities}), we obtain
\beqa
b(x)&=&{[l_1(x-{i\ga\over{2}})+l_2(x-{i\ga\over{2}})]l_2(x+{i\ga\over{2}})\over{
l_1(x-{i\ga\over{2}})l_1(x+{i\ga\over{2}})}} \label{bT1Q} \\
&=&{T_1(x-{i\ga\over{2}})\over{\sinh(2x+2i\ga)}}{\phi(x-{i\ga\over{2}})\over{
\phi(x+{i\ga\over{2}})\phi(x+{3i\ga\over{2}})}}
{B^{(+)}(x+{i\ga\over{2}})\over{B^{(-)}(x-{i\ga\over{2}})B^{(-)}(x+{i\ga\over{2}})}}
{Q(x+{3i\ga\over{2}})\over{Q(x-{3i\ga\over{2}})}} \,. \non
\eeqa
In terms of the quantities (\ref{newqdef}), (\ref{checkTdef}), we rewrite this relation 
in the compact form
\be
b(x)=C_b(x)\, \check T_1(x-{i\ga\over{2}})
{q(x+{3i\ga\over{2}}-{i \pi\over 2})\over{q(x-{3i\ga\over{2}}+{i \pi\over 2})}} \,,
\label{simplebandbarbT1Q}
\ee
with
\be
C_b(x) &=&{\mu(x-{i\ga\over{2}}) \phi(x-{i\ga\over{2}})\over{\sinh(2x+2i\ga)
\phi(x+{i\ga\over{2}})\phi(x+{3i\ga\over{2}})}}
{B^{(+)}(x+{i\ga\over{2}})\over{B^{(-)}(x-{i\ga\over{2}})B^{(-)}(x+{i\ga\over{2}})}} 
\,. \label{Cb}
\ee

Defining
\beq
B(x)=1+b(x) \,,\qquad {\bar B}(x)=1+{\bar b}(x) = B(-x) \,,
\eeq
we also obtain
\beq
T_2(x)=\la_3(x)\, B(x)=\la_1(x)\, {\bar B}(x) \,.
\label{T2B}
\eeq

In terms of the quantities (\ref{newqdef}), (\ref{checkTdef}), we can reexpress (\ref{T2B}) as 
\be
\check{T}_2(x)
&=&t_{-}(x)\,
{q(x-{3i\ga\over{2}}+{i\pi\over{2}})\over{q(x+{i\ga\over{2}}-{i\pi\over{2}})}}B(x) \,,
\label{T2Bq} \\
&=&t_{+}(x)\,
{q(x+{3i\ga\over{2}}-{i\pi\over{2}})\over{q(x-{i\ga\over{2}}+{i\pi\over{2}})}}{\bar B}(x) \,,
\label{T2BarBq}
\ee
with
\beq
t_{\pm}(x)={\sinh(2x\mp 2i\ga)\over \mu(x)}
B^{(\pm)}(x-{i\ga\over{2}})B^{(\pm)}(x+{i\ga\over{2}})
\phi(x\mp{3i\ga\over{2}})\phi(x\mp{i\ga\over{2}}).
\eeq

We note that $B(x)$ has zeros just above the real axis, and $\bar
B(x)$ has zeros just below the real axis.  Indeed, due to the factor
$Q(x+ \frac{i\ga}{2})$ in the definition of $\la_{3}(x)$ (\ref{T2}),
and the fact that the imaginary parts of the Bethe roots have
magnitude $> \frac{\ga}{2}$, we see that $\la_{3}(x)$ has poles just
above the real axis (and also just below the line $\Im m\, x =-\ga$).
It follows from (\ref{T2B}) that $B(x)$ must have corresponding zeros
to cancel these poles, since the product $\la_3(x)B(x) = T_2(x)$ is
analytic.  Similarly, since $\la_{1}(x)$ has poles just below the real
axis (and also just above the line $\Im m\, x = \ga$), $\bar B(x)$ has
corresponding zeros to cancel these poles.

Finally, we define the auxiliary functions $y(x)$ and $Y(x)$,
\be
y(x)={T_0(x)\, T_2(x)\over f(x)} \,, \qquad Y(x)=1+y(x) \,,
\label{yy}
\ee 
in terms of which the fusion relation (\ref{T1T1}) becomes 
\beq
\check T_1(x-{i\ga\over{2}})\, \check T_1(x+{i\ga\over{2}})=
{f(x)\, Y(x)\over \mu(x-{i\ga\over{2}})\,  \mu(x+{i\ga\over{2}})} \,.
\label{T1T1Y}
\eeq

\subsection{Derivation of NLIE}\label{subsec:NLIEspin1}


Since $\ln \check{T}_2(x)$ is analytic near the real axis, Cauchy's theorem gives
\beq
0=\oint_C dx\ [\ln\check{T}_2(x)]'' e^{ikx} = 
\int_{C_1}[\ln\check{T}_2(x)]'' e^{ikx} dx + 
\int_{C_2}[\ln\check{T}_2(x)]'' e^{ikx} dx \,,
\label{cauchy1}
\eeq
where we choose the contour $C$ as in Figure \ref{fig:contour}, with 
$\epsilon$ such that max $\{ y_{k}\} - \frac{\ga}{2}  < \epsilon$. 
These integrals can be written using (\ref{T2Bq}) and (\ref{T2BarBq}) as
\be
\int_{C_1}[\ln\check{T}_2(x)]'' e^{ikx} dx 
&=&\int_{C_1} dx\ \left\{ \ln t_{-}(x) \right\}'' e^{ikx}+
\int_{C_1} dx\ \left\{ \ln \left[{q(x-\frac{3i\ga}{2} + \frac{i \pi}{2})\over 
q(x+\frac{i\ga}{2} - \frac{i \pi}{2})}\right] \right\}'' e^{ikx} \non \\
&+&\int_{C_1} dx\ \left\{ \ln B(x) \right\}'' e^{ikx} \,, \label{cauchy1a} \\
\int_{C_2}[\ln\check{T}_2(x)]'' e^{ikx} dx
&=& \int_{C_2} dx\ \left\{ \ln t_{+}(x) \right\}'' e^{ikx}+
\int_{C_2} dx\ \left\{ \ln\left[{q(x+\frac{3i\ga}{2} - \frac{i \pi}{2})\over 
q(x-\frac{i\ga}{2} + \frac{i \pi}{2})}\right] 
\right\}'' e^{ikx} \non \\ 
&+&\int_{C_2} dx\ \left\{ \ln \bar B(x)\right\}'' e^{ikx} \,,
\label{cauchy1b}
\ee
so that the quantities $q(x+\frac{i\ga}{2} - \frac{i \pi}{2})$ and 
$q(x-\frac{i\ga}{2} + \frac{i \pi}{2})$
are integrated along 
$C_{1}$ and $C_{2}$, respectively, and not the other way around.
\footnote{Because of our choice of $\epsilon$, the quantities
$q(x+\frac{i\ga}{2} - \frac{i \pi}{2})$ and $B(x)$ are nonzero along
$C_{1}$, and $q(x-\frac{i\ga}{2} + \frac{i \pi}{2})$ and $\bar B(x)$
are nonzero along $C_{2}$.} 

In terms of Fourier transforms defined by (\ref{fouriertransfdef}),
we can rewrite (\ref{cauchy1a}) and (\ref{cauchy1b}) as 
\footnote{Note that all the integrals of $q$ have been expressed in terms 
of $\widehat{Lq''}(k)$. This would not have been possible if we had 
interchanged $C_{1}$ and $C_{2}$ in (\ref{cauchy1a}) and (\ref{cauchy1b}).}
\be
\widehat{{\cal L}\check{T}''_2}(k) &=&
\widehat{{\cal L}t''_-}(k)+
\left[e^{\left({\ga\over{2}}-{\pi\over{2}}\right)k}
-e^{\left({\pi\over{2}}-{3\ga\over{2}}\right)k}\right]
\widehat{Lq''}(k)+\widehat{{\cal L}B''}(k) \,,
\label{LT21} \\
\widehat{L\check{T}''_2}(k) &=&
\widehat{Lt''_+}(k)+
\left[e^{\left({3\ga\over{2}}-{\pi\over{2}}\right)k}
-e^{\left({\pi\over{2}}-{\ga\over{2}}\right)k}\right]\widehat{Lq''}(k)+
\widehat{L{\bar B}''}(k) \,,
\label{LT22}
\ee
respectively.
In obtaining these results, we use the periodicity of $q(x)$ as in 
(\ref{Qperiodicity})  to make the imaginary part of the argument negative.

Adding (\ref{LT21}) and (\ref{LT22}), and remembering the fact (\ref{cauchy1})
\be
\widehat{{\cal L}\check{T}''_2}(k) + \widehat{L\check{T}''_2}(k) = 0 
\,, \label{cauchyagain}
\ee 
we obtain an expression for $\widehat{Lq''}(k)$,
\beq
\left[e^{\left({\pi\over{2}}-{3\ga\over{2}}\right)k}
-e^{\left({\ga\over{2}}-{\pi\over{2}}\right)k}
-e^{\left({3\ga\over{2}}-{\pi\over{2}}\right)k}
+e^{\left({\pi\over{2}}-{\ga\over{2}}\right)k}\right]\widehat{Lq''}(k)=
\widehat{L{\bar B}''}(k)+\widehat{{\cal L}B''}(k)+D_q(k) \,,
\label{LqBB}
\eeq
where we define
\be
D_q(k)=\widehat{Lt''_+}(k)+\widehat{{\cal L}t''_-}(k) \,.
\label{Dqdef}
\ee

From the expression (\ref{simplebandbarbT1Q}) for $b(x)$ 
and the fact that $\check T_{1}(x)$ is ANZ near the real axis, we obtain
\beqa
\widehat{Lb''}(k)&=&e^{-\frac{\ga k}{2}}\widehat{L\check{T}''_1}(k)+\widehat{Lq''}(k)
\left[e^{\left({3\ga\over{2}}-{\pi\over{2}}\right)k}-
e^{\left({\pi\over{2}}-{3\ga\over{2}}\right)k}\right]+\widehat{LC''_b}(k) \,.
\label{LbLT1Lq}
\eeqa

From the fusion relation (\ref{T1T1Y}), we obtain
\beq
\left(e^{{\ga k\over{2}}}+e^{-{\ga k\over{2}}}\right)\widehat{L\check{T}''_1}(k)
=\widehat{LY''}(k) + \widehat{Lf''}(k) - \left(e^{\ga k\over 
2}+e^{-{\ga k\over 2}}\right) \widehat{L\mu''}(k) + e^{\ga k\over 2} \delta(k)
\,,
\label{LT1Y}
\eeq
where $\delta(k)$ is again given by (\ref{deltaresult}).
Substituting Eqs.(\ref{LqBB}) and (\ref{LT1Y}) into Eq.(\ref{LbLT1Lq}),
we obtain
\beqa
\widehat{Lb''}(k)&=&{e^{\left({3\ga\over{2}}-{\pi\over{2}}\right)k}-
e^{\left({\pi\over{2}}-{3\ga\over{2}}\right)k}\over{
e^{\left({\pi\over{2}}-{3\ga\over{2}}\right)k}
-e^{\left({\ga\over{2}}-{\pi\over{2}}\right)k}
-e^{\left({3\ga\over{2}}-{\pi\over{2}}\right)k}
+e^{\left({\pi\over{2}}-{\ga\over{2}}\right)k}}}\left[
\widehat{L{\bar B}''}(k) +\widehat{{\cal L}B''}(k)+D_q(k)\right] \non \\
&+& {e^{-{\ga k\over{2}}}\over{e^{{\ga k\over{2}}}+e^{-{\ga k\over{2}}}}}
\left[\widehat{LY''}(k)+\widehat{Lf''}(k)\right] +\widehat{LC''_b}(k)
- e^{-{\ga k\over 2}}\widehat{L\mu''}(k)
+{\delta(k)\over e^{\ga k\over 2}+e^{-{\ga k\over 2}}} \non \\
&=&-\widehat G(k)\left[\widehat{L{\bar B}''}(k)+\widehat{{\cal L}B''}(k)\right]+
\widehat G_2(k)\, \widehat{LY''}(k)+ C(k) \,,
\label{Lbnlie}
\eeqa
where
\beqa
\widehat G(k)&=&
{\sinh\left((\pi-3\ga)\frac{k}{2}\right)\over
2\cosh{\ga k\over{2}}\sinh\left((\pi-2\ga)\frac{k}{2}\right)} \,, 
\label{Gspin1} \\
\widehat G_2(k)&=&{e^{-{\ga k\over{2}}}\over{e^{{\ga 
k\over{2}}}+e^{-{\ga k\over{2}}}}} \,, \label{G2def} \\
C(k)&=&-\widehat G(k)\, D_q(k)+ 
\widehat G_2(k)\, \widehat{Lf''}(k) +\widehat{LC''_b}(k)
- e^{-{\ga k\over 2}}\widehat{L\mu''}(k)
+{\delta(k)\over e^{\ga k\over 2}+e^{-{\ga k\over 2}}} \,. 
\label{spin1Cdef}
\eeqa
We find that $C(k)$ is given by (\ref{Ckfinal})
\beqa
C(k) &=& 2\pi k \Bigg\{
N\left({e^{i\La k}+e^{-i\La k}\over{2\cosh{\ga k\over{2}}}}\right)
+ {\left[\sinh\left((\eta_{+}-{\pi\over{2}})k\right)+
\sinh\left((\eta_{-}-{\pi\over{2}})k\right)\right]\over{
2\cosh{\ga k\over{2}}\sinh\left(({\pi\over{2}}-\ga)k\right)}}\nonumber\\
&+&{\cosh \frac{\ga k}{4} \sinh\left((3\ga -\pi)\frac{k}{4}\right)\over
\cosh \frac{\ga k}{2}\sinh\left((2\ga -\pi)\frac{k}{4}\right)}
\Bigg\} \,.
\label{Ckfinalagain}
\eeqa


We now turn to the third and final NLIE equation. From the definitions of $y(x)$ (\ref{yy}) 
and $\check T_{2}(x)$ (\ref{checkTdef}), we obtain
\beq
\widehat{L y''}(k)=\widehat{L\check{T}''_2}(k) + \widehat{LT''_0}(k) 
+ \widehat{L\mu''}(k) -\widehat{Lf''}(k) \,.
\label{ftLy}
\eeq
To evaluate $\widehat{L\check{T}''_2}(k)$, we combine Eqs. 
(\ref{LT21}) and (\ref{LT22}) to cancel $\widehat{Lq''}(k)$, namely,
\beq
-e^{{\ga k\over{2}}}\widehat{{\cal L}\check{T}''_2}(k)+e^{-{\ga k\over{2}}}
\widehat{L\check{T}''_2}(k) \,, \non 
\eeq
which together with (\ref{cauchyagain}) gives
\beq
\left(e^{{\ga k\over{2}}}+e^{-{\ga k\over{2}}}\right)
\widehat{L\check{T}''_2}(k)=e^{-{\ga k\over{2}}}\widehat{L{\bar B}''}(k)-
e^{{\ga k\over{2}}}\widehat{{\cal L}B''}(k)+D_T(k) \,,
\label{ftT2result}
\eeq
where we define
\be
D_T(k) = e^{-{\ga k\over{2}}}\widehat{Lt''_+}(k) -
e^{{\ga k\over{2}}}\widehat{{\cal L}t''_-}(k) \,.
\label{DTdef}
\ee
Substituting the result (\ref{ftT2result}) for $\widehat{L\check{T}''_2}(k)$ into (\ref{ftLy}), we obtain
\be
\widehat{Ly''}(k) = -\widehat G_2(-k)\, \widehat{{\cal L}B''}(k)+
\widehat G_2(k)\, \widehat{L{\bar B}''}(k)+ C_y(k) \,,
\ee
where we define
\be
C_y(k)={D_T(k)\over{e^{{\ga k\over{2}}}+e^{-{\ga k\over{2}}}}}
+ \widehat{LT''_0}(k) + \widehat{L\mu''}(k)-\widehat{Lf''}(k) \,.
\label{Cydef}
\ee
We find (\ref{Cyfinal})
\be
C_y(k)= 4\pi k \widehat G_2(-k) \,. \label{Cyresult}
\ee

In summary, the NLIEs of the lattice SSG model with Dirichlet boundary conditions in 
Fourier space are 
\beqa
\widehat{Lb''}(k)&=&-\widehat G(k)\left[\widehat{L{\bar 
B}''}(k)+\widehat{{\cal L}B''}(k)\right]+
\widehat G_2(k)\, \widehat{LY''}(k)+ C(k) \,, \label{NLIEbk}\\
\widehat{Ly''}(k)&=&\widehat G_2(k)\, \widehat{L{\bar B}''}(k)-
\widehat G_2(-k)\, \widehat{{\cal L}B''}(k)+ C_y(k)
\,, \label{NLIEyk}
\eeqa
where $C(k)$ and $C_y(k)$ are given by Eqs. (\ref{Ckfinalagain}) and 
(\ref{Cyresult}), respectively.
Passing to coordinate space, integrating twice, and taking the 
continuum limit, we obtain 
\be
\ln \bff(\theta) &=& 
\int_{-\infty}^{\infty}d\theta'\ \Gf(\theta-\theta'-i\varepsilon) 
\ln \Bf(\theta'+i\varepsilon) 
-\int_{-\infty}^{\infty}d\theta'\ \Gf(\theta-\theta'+i\varepsilon) \ln \bar 
\Bf(\theta'-i\varepsilon)\non \\
&+& \int_{-\infty}^{\infty}d\theta'\ \Gf_{2}(\theta-\theta'+i\varepsilon) \ln 
\Yf(\theta'-i\varepsilon) + i 2mL \sinh \theta + i\,  
\Pf_{bdry}(\theta) -i\pi\,, 
\non \\
\ln \bar \bff(\theta) &=& 
-\int_{-\infty}^{\infty}d\theta'\ \Gf(\theta-\theta'-i\varepsilon) 
\ln \Bf(\theta'+i\varepsilon) 
+\int_{-\infty}^{\infty}d\theta'\ \Gf(\theta-\theta'+i\varepsilon) \ln \bar 
\Bf(\theta'-i\varepsilon)\non \\
&+& \int_{-\infty}^{\infty}d\theta'\ \Gf_{2}(\theta'-\theta+i\varepsilon) \ln 
\Yf(\theta'+i\varepsilon) - i 2mL \sinh \theta - i\,  
\Pf_{bdry}(\theta) + i\pi \,, \non \\
\ln \yf(\theta) &=& \int_{-\infty}^{\infty}d\theta'\ 
\Gf_{2}(\theta-\theta'+i\varepsilon) \ln \bar 
\Bf(\theta'-i\varepsilon) + \int_{-\infty}^{\infty}d\theta'\ 
\Gf_{2}(\theta'-\theta+i\varepsilon) \ln 
\Bf(\theta'+i\varepsilon)  \non \\
&+& i\, \Pf_{y}(\theta) \,. 
\label{NLIEspin1coord}
\ee
As in the spin-1/2 case, the continuum limit consists of taking $\La
\rightarrow \infty$, $N \rightarrow \infty$ and lattice spacing $a
\rightarrow 0$, such that the interval length $L\equiv x_{+}-x_{-}$
and the soliton mass $m$ are given by (\ref{continuumlimit}).  The
renormalized rapidity $\theta$ is again given by
(\ref{renormrapidity}), and we have defined $\bff(\theta)=b(\frac{\ga
\theta}{\pi})$, etc.  as in (\ref{mathfrakdefs}). Hence, the kernel
$\Gf(\theta)$ is given by
\be
\Gf(\theta) =  {\ga\over 2\pi^{2}} \int_{-\infty}^{\infty}dk\ e^{-i 
k \ga \theta/\pi}\ \widehat G(k) \,,
\ee
where 
$\widehat G(k)$ is given by (\ref{Gspin1}); and $\Gf_{2}(\theta)$ is 
defined similarly in terms of $\widehat G_{2}(k)$ (\ref{G2def}),
\be
\Gf_{2}(\theta) =  {\ga\over 2\pi^{2}} \int_{-\infty}^{\infty}dk\ e^{-i 
k \ga \theta/\pi}\ \widehat G_{2}(k) 
= {i\over 2\pi \sinh \theta} \,,
\label{G2theta}
\ee
where $\theta$ has a slightly positive imaginary part.
Finally, $\Pf_{bdry}(\theta)$ is given by
\be
\Pf_{bdry}(\theta) = \int_{0}^{\frac{\ga \theta}{\pi}}dx'\, R(x') = \frac{1}{2} 
\int_{-\frac{\ga \theta}{\pi}}^{\frac{\ga \theta}{\pi}}dx'\, R(x')
=  {\ga\over 4\pi^{2}} \int_{-\theta}^{\theta} d\theta' \int_{-\infty}^{\infty}dk\ 
e^{-i k \ga \theta'/\pi}\ \hat R(k)\,,
\ee
where $\hat R(k)$ is given by
\be
\hat R(k) = 2 \pi \Bigg\{ 
{\left[\sinh\left((\eta_{+}-{\pi\over{2}})k\right)+
\sinh\left((\eta_{-}-{\pi\over{2}})k\right)\right]\over{
2\cosh{\ga k\over{2}}\sinh\left((\pi-2\ga)\frac{k}{2}\right)}}
+{\cosh \frac{\ga k}{4} \sinh\left((3\ga -\pi)\frac{k}{4}\right)\over
\cosh \frac{\ga k}{2}\sinh\left((2\ga -\pi)\frac{k}{4}\right)}
\Bigg\} \,; \label{FTPbdry}
\ee 
and $\Pf_{y}(\theta)$ is given by
\be
\Pf_{y}(\theta) = 4\pi \int_{-\infty}^{\theta}d\theta'\ 
\Gf_{2}(-\theta') = -2i \ln 
\tanh \frac{\theta}{2} - 2\pi \,,
\label{Pytheta}
\ee 
where $\theta$ has a slightly negative imaginary part.
The integration constants are explained in Section \ref{sec:spin1integconst}.

These NLIE equations are similar to those for the periodic 
chain \cite{Du, Su2, HRS}, with additional boundary terms involving
$\Pf_{bdry}(\theta)$ or $\Pf_{y}(\theta)$. These terms 
constitute one of our main results. As we shall see, these boundary 
terms make essential contributions to the boundary $S$ matrix (IR limit) and to 
the effective central charge (UV limit).

\subsection{Vacuum and Casimir energies}\label{subsec:spin1vacuumCasimir}

The energy computation is similar to the one for the spin-1/2 case in 
Section \ref{subsec:spinhalfvacuumCasimir}. We see from (\ref{yy})
that $T_{2}(x)$ can be expressed in terms of $y(x)$, 
\be
T_2(x)={f(x)\, y(x)\over T_0(x)} \,.
\ee 
It follows from the energy definition (\ref{energydef2}) together with the 
fact (\ref{FTfact}) that
\be
E=-\frac{g}{a} \int \frac{dk}{2\pi} 
\left[e^{-ik(\Lambda + \frac{i\ga}{2})}-e^{-ik(\Lambda - 
\frac{i\ga}{2})}\right] \left[\widehat{Lf'}(k) +  
\widehat{Ly'}(k) - \widehat{LT'_{0}}(k) \right] \,.
\ee 
Recalling the results for $\widehat{Lf''}(k)$ (\ref{ftfresult}) and 
$\widehat{Ly''}(k)$ (\ref{NLIEyk}), we obtain
\be
E &=&-\frac{g}{a}\int \frac{dk}{2\pi} e^{-ik\Lambda}\Bigg\{
\widehat{L\phi'}(k) (e^{\ga k}-1)(e^{(\ga-\pi)k}+e^{-2\ga k}) \non\\
&+& 2\sinh \left({\ga k\over{2}}\right)\Big[e^{-{\ga k\over{2}}}\widehat{LB^{(+)'}}(k)+
e^{{\ga k\over{2}}}\widehat{LB^{(-)'}}(k) + 4\pi i \widehat G_{2}(-k) \non \\
&+& 2\pi \frac{\psi_{2}(k)}{(-ik)}\left(e^{-\ga k}+e^{(\ga - 
\frac{\pi}{2})k}-1\right) \Big] 
+\frac{1}{\cosh{\ga k\over{2}}}\left[\widehat{L{\bar B}'}(k)
+\widehat{{\cal L}B'}(k)\right] \Bigg\} \,.
\ee

The first term gives the bulk vacuum energy, which can be written explicitly as
\beq
E_{B}=\frac{2Ng}{i a}\int_{-\infty}^{\infty} dk e^{-2i\Lambda k}
{\sinh \frac{\ga k}{2}\cosh\left((\frac{3\ga}{2}-\frac{\pi}{2})k\right)
\over \sinh\frac{\pi k}{2}} \,.
\label{FTBenergy1}
\eeq
The second term gives (with the help of (\ref{Bpmk})) the boundary 
vacuum energy
\be
E_{b}&=&\frac{2g}{i a}\int_{-\infty}^{\infty} dk e^{-i\Lambda k}
\sinh\left({\ga k\over{2}}\right)\Bigg\{\frac{1}{\sinh{\pi k\over 2}}
\left[\cosh\left((\eta_{+}-\frac{\ga}{2}-\frac{\pi}{2})k \right)
+(\eta_{+}\to \eta_{-})\right] \non \\
&+& {e^{\frac{\ga k}{2}}\over \cosh \frac{\ga k}{2}}
+ \frac{e^{(\frac{\pi}{4}-\ga )k}
+e^{(\ga - \frac{\pi}{4})k}-e^{\frac{\pi k}{4}}}{2\sinh{\pi k\over 4}}
\Bigg\}\,,
\label{FTbenergy1}
\ee
and the third term gives the Casimir energy, 
\beq
E_C=-\frac{g}{a}\int_{-\infty}^{\infty} \frac{dk}{2\pi} e^{-i\Lambda k}
\frac{1}{\cosh{\ga k\over{2}}}\left[
\widehat{L{\bar B}'}(k)+\widehat{{\cal L}B'}(k)\right] \,.
\label{FTcasimir1}
\eeq

In order to take the continuum limit, we adopt (as in the spin-1/2 case) 
the renormalization procedure of
keeping only the (finite) terms that can be expressed in terms of the
physical mass $m$ (\ref{continuumlimit}).  We implement this procedure
by closing the integral contours in the lower half plane, and
selecting only the contribution from the residue at
$k=-\frac{i\pi}{\ga}$.  Since the integrand in (\ref{FTBenergy1})  is
analytic at $k=-\frac{i\pi}{\ga}$, the bulk energy vanishes,
\be
E_{B}=0 \,, \label{SSGbulkvacuumenergy}
\ee
in agreement with known results (see, e.g., the second reference in
\cite{Du}).  We obtain from (\ref{FTbenergy1}) that the boundary vacuum
energy does {\it not} vanish, however, and is given by
\be
E_{b}=m \,. \label{SSGboundvacuumenergy}
\ee
That is, each boundary contributes the vacuum energy $m/2$.
Note that this result is independent of the boundary parameters. We 
shall present further support for this result in Section \ref{sec:intermediate}.
Finally, we find that the Casimir energy
(\ref{FTcasimir1}) is given by
\beq
E_C=\frac{m}{2\pi}\Im m\int_{-\infty}^{\infty} d\th\
\sinh(\th-i\varepsilon)\ln \bar \Bf(\th-i\varepsilon) \,. 
\label{Casimirspin1}
\eeq

\section{Infrared limit}\label{sec:irlimit}

In this Section we analyze the IR limit $m L \rightarrow \infty$.  In
particular, we compute the SSG soliton boundary $S$ matrix (\ref{IRresult}), 
and show that
it coincides with one proposed by Bajnok {\it et al.} \cite{BPT}.  In
making this identification, we determine the ``lattice
$\leftrightarrow$ IR'' relation for the boundary SSG parameters 
(\ref{boundlatIR}).

Before starting this computation, we note the relations among 
the three {\it bulk} SSG parameters (see 
Figure \ref{fig:parameters}). Let $\lambda$ denote the IR bulk SSG 
parameter, and recall that $\beta$ and $\ga$ are our UV and lattice bulk SSG parameters, 
respectively. (See Eqs. (\ref{ssg}),  (\ref{bulkhamiltonian}).)
The SSG soliton bulk
$S$ matrix \cite{ABL} has the product form $SG(\theta\,; \lambda) 
\otimes RSOS(\theta)$, 
where $SG(\theta\,; \lambda) $ is the SG soliton bulk $S$ matrix 
\cite{ZZ}, and $RSOS(\theta)$ is the quantum-group restricted 
SG model bulk $S$ matrix \cite{RSOS}. The bulk ``UV
$\leftrightarrow$ IR'' relation is given by \cite{ABL},
\be
\frac{2\pi}{\beta^{2}}-\frac{1}{2}= \lambda \,.
\label{bulkUVIR}
\ee
The bulk ``lattice $\leftrightarrow$ IR'' relation is given by
\be
\frac{1}{\frac{\pi}{\ga}-2} = \lambda \,.
\label{bulklatIR}
\ee 
This relation can be readily inferred from a comparison of the kernel
$\widehat{G}(k)$ (\ref{Gspin1}) with the integral representation of
the SG soliton-soliton scattering amplitude $SG_{++}(\theta\,; \lambda)$,
\be
\frac{1}{i}\frac{d}{d\theta}\ln SG_{++}(\theta\,; \lambda)
=\int_{-\infty}^{\infty}dk\, e^{-i k \theta} \,
{\sinh\left((\frac{1}{\lambda}-1)\frac{\pi k}{2}\right)\over
2 \cosh\frac{\pi k}{2} \sinh \frac{\pi k}{2\lambda}} \,.
\ee 
The relations (\ref{bulkUVIR}), (\ref{bulklatIR}) imply the
bulk ``UV $\leftrightarrow$ lattice'' relation
\be
\beta^{2}=4\left(\pi-2\ga\right) \,.
\label{bulkUVlat}
\ee
The condition $0 < \beta^{2} < 4\pi$ therefore implies $0 < \ga < 
\frac{\pi}{2}$. The free Fermion point $\lambda=1$ corresponds to 
$\ga = \frac{\pi}{3}$. In the ``repulsive'' domain 
$\ga \in (0\,, \frac{\pi}{3})$, the 
particle spectrum consists only of supersymmetric multiplets of 
solitons and antisolitons. In the ``attractive'' domain
$\ga \in (\frac{\pi}{3}\,, \frac{\pi}{2})$, the 
spectrum also includes bound states, namely, supersymmetric 
multiplets of breathers of mass \cite{ABL}
\be
m_{n} = 2m \sin \left(\frac{n \pi}{2\lambda}\right) \,, \qquad n = 
1\,, 2\,, \ldots \,, \lfloor \lambda \rfloor \,.
\label{breathermass}
\ee 

We turn now to the boundary theory.  Let $\xi_{\pm}$ denote the IR
boundary SSG parameters, and recall that $\varphi_{\pm}$ and
$\eta_{\pm}$ are our UV and lattice boundary SSG parameters,
respectively.  (See Eqs.  (\ref{Dirichlet}), (\ref{diagbt}).)

We define the boundary $S$ matrices (reflection factors)
$R(\theta_{h}\,; \lambda \,, \xi_{\pm})$ for a soliton with mass $m$
and rapidity $\theta_{h}$ by the Yang equation for a single soliton on an interval
of length $L \gg 1/m$,
\be
e^{i 2mL \sinh \theta_{h}} R(\theta_{h}\,; \lambda\,, \xi_{-})\, 
R(\theta_{h}\,; \lambda\,, \xi_{+}) 
= 1 \,. \label{yang}
\ee
We shall compute these $S$ matrices by deriving a similar relation
from the IR limit $m L \rightarrow \infty$ of the NLIE for a state of
one hole with rapidity $\theta_{h}$,
\be
\ln \bff(\theta) &=&  i 2mL \sinh \theta + i \Pf_{bdry}(\theta) 
+ i \chi(\theta-\theta_{h}) + i \chi(\theta+\theta_{h}) \non \\
&+& \int_{-\infty}^{\infty}d\theta'\ \Gf_{2}(\theta-\theta'+i\varepsilon) \ln 
\Yf(\theta'-i\varepsilon) -i\pi \,, 
\label{NLIEIRb} \\
\ln \yf(\theta) &=& i \Pf_{y}(\theta) + 
i g_{y}(\theta-\theta_{h})
+ i g_{y}(\theta+\theta_{h}) 
\,,
\label{NLIEIRy}
\ee
where $\chi(\theta)$ and $g_{y}(\theta)$ are the hole source terms 
\cite{Su2, HRS}
\be
\chi(\theta) = 2\pi \int_{0}^{\theta}d\theta'\, \Gf(\theta') \,, 
\qquad 
g_{y}(\theta) = -i \ln \tanh \frac{\theta}{2} + \frac{\pi}{2}
\,, \label{sourceterms}
\ee 
and where, in the latter equation, $\theta$ has a slightly negative 
imaginary part.
Indeed, since $\ln \bff(\theta_{h})$ is $i\pi$ times an odd integer
(see, e.g., \cite{Su2,HRS}), 
evaluating Eq. (\ref{NLIEIRb}) at $\theta_{h}$ and  
exponentiating both sides gives
\be
e^{i 2mL \sinh \theta_{h}}\, e^{i\Pf_{bdry}(\theta_{h}) +i 
\chi(2\theta_{h})+{\cal K}(\theta_{h})} = 1 \,, \label{NLIEyang}
\ee 
where ${\cal K}(\theta)$ is the convolution term in (\ref{NLIEIRb}),
\be
{\cal K}(\theta) \equiv 
\int_{-\infty}^{\infty}d\theta'\ \Gf_{2}(\theta-\theta'+i\varepsilon) 
\ln \Yf(\theta'-i\varepsilon) \,. \label{calKdef}
\ee 
Comparing (\ref{NLIEyang}) with the Yang equation (\ref{yang}),
we conclude that the product of boundary $S$ matrices is given by
\be
R(\theta_{h}\,; \lambda\,, \xi_{-})\, 
R(\theta_{h}\,; \lambda\,, \xi_{+})  = e^{i\Pf_{bdry}(\theta_{h}) +i 
\chi(2\theta_{h})+{\cal K}(\theta_{h})} \,. \label{prodRR}
\ee

We evaluate first the factor $e^{{\cal K}(\theta_{h})}$ which, as we 
shall see, is the RSOS factor. To this end, 
we observe from (\ref{Pytheta}), (\ref{NLIEIRy}) and (\ref{sourceterms})
that $\yf(\theta)$ is given by 
\be
\yf(\theta) = - \tanh^{2}\frac{\theta}{2} \tanh \frac{1}{2}(\theta-\theta_{h})
\tanh \frac{1}{2}(\theta+\theta_{h})\,.
\ee
It follows that
\be
\ln \Yf(\theta) = \ln (1+\yf(\theta)) = \ln \left[
{\cosh^{2} \frac {\theta_{h}}{2} \cosh \theta\over
\cosh^{2} \frac {\theta}{2} \cosh\frac{1}{2}(\theta-\theta_{h}) 
\cosh\frac{1}{2}(\theta+\theta_{h})} \right] \,.
\ee 
The convolution term (\ref{calKdef}) is therefore given by
\be
{\cal K}(\theta) = \ln \cosh \frac{\theta_{h}}{2} + Q_{1}(\theta) - 2Q_{2}(\theta) 
-Q_{2}(\theta-\theta_{h}) -Q_{2}(\theta+\theta_{h}) \,,
\label{calK}
\ee
where \cite{HRS}
\be
Q_{1}(\theta) &=& \int_{-\infty}^{\infty}d\theta'\ \Gf_{2}(\theta-\theta'+i\varepsilon) 
\ln \cosh (\theta' -i\varepsilon) = -\frac{i}{2} \tan^{-1}\sinh \theta 
+ \frac{1}{2} \ln \cosh \theta \,, \non \\ 
Q_{2}(\theta) &=& \int_{-\infty}^{\infty}d\theta'\ \Gf_{2}(\theta-\theta'+i\varepsilon) 
\ln \cosh  \frac{1}{2}(\theta' -i\varepsilon) = -\frac{i}{2} \chi_{2} (\theta)
+ \frac{1}{2} \ln \cosh  \frac{\theta}{2} \,,
\ee
with
\be
\chi'_{2} (\theta) = \int_{-\infty}^{\infty} dk\, 
e^{-i k \theta} {1\over 4 \cosh^{2} \frac{\pi k}{2}} \,, \qquad 
\chi_{2}(0) = 1\,.
\label{chi2}
\ee 
It follows that
\be
{\cal K}(\theta_{h}) = \frac{i}{2}\left[ \chi_{2}(2\theta_{h}) + 
2 \chi_{2}(\theta_{h}) -  \tan^{-1}\sinh \theta_{h} \right] \,.
\ee
Differentiating with respect to $\theta_{h}$, and then making use of 
(\ref{chi2}) and the Fourier transform result
\be
{1\over  \cosh\theta} = \int_{-\infty}^{\infty} dk\, 
e^{-i k \theta} {1\over 2\cosh \frac{\pi k}{2}}\,,
\ee 
we obtain
\be
{d\over d\theta_{h}} {\cal K}(\theta_{h}) =
\frac{i}{2}\left[ 2\chi'_{2}(2\theta_{h}) + 
2 \chi'_{2}(\theta_{h}) -  {1\over  \cosh\theta_{h}} \right]
= \frac{i}{4}\int_{-\infty}^{\infty} dk\, {e^{-2 i k \theta_{h}}\over
\cosh^{2} \frac{\pi k}{2} \cosh^{2} \pi k} \,.
\ee
Upon integrating, we conclude that \footnote{We use $\sim$ to 
denote equality up to crossing factors of the form $e^{const\ 
\theta}$. Such factors have also not been obtained in the bulk case 
\cite{HRS}.}
\be
e^{{\cal K}(\theta_{h})} \sim P_{min}(\theta_{h})^{2} \,,
\label{calKresult}
\ee 
where $P_{min}(\theta)$ is a reflection factor of the boundary tricritical 
Ising model \cite{AK}, whose integral representation is given by 
\cite{AN2}
\be
P_{min}(\theta) \sim \exp \left\{ \frac{i}{8}\int_{0}^{\infty} {dt\over t}\, 
{\sin (2 t \theta/\pi)\over \cosh^{2}\frac{t}{2} \cosh^{2}t}
\right\} \,. 
\label{Pmin}
\ee
We stress that the boundary term $ \Pf_{y}(\theta)$ in (\ref{NLIEIRy}) is essential 
for obtaining this result. Since $ \Pf_{y}(\theta)$ is independent of 
the boundary parameters, so is the result (\ref{calKresult}).

We turn now to the remaining factor $e^{i\Pf_{bdry}(\theta_{h}) +i 
\chi(2\theta_{h})}$ in (\ref{prodRR}) which, as we shall see, is the 
sine-Gordon factor. Differentiating with respect to $\theta_{h}$,
and recalling the Fourier transform results (\ref{Gspin1}), 
(\ref{FTPbdry}), we obtain
\be
\lefteqn{{d\over d\theta_{h}}\left[\Pf_{bdry}(\theta_{h}) + 
\chi(2\theta_{h})\right] = \Pf'_{bdry}(\theta_{h}) + 
2\chi'(2\theta_{h})}  \non \\
& & = \int_{-\infty}^{\infty} dk\, 
e^{-i k \theta_{h}} \Bigg\{
{\left[\sinh\left((\eta_{+}-{\pi\over{2}})\frac{\pi k}{\ga}\right)+
\sinh\left((\eta_{-}-{\pi\over{2}})\frac{\pi k}{\ga}\right)\right]\over{
2\cosh{\pi k\over{2}}\sinh\left((\frac{\pi}{\ga}-2)\frac{\pi 
k}{2}\right)}} \non \\
& & \qquad +2{\sinh \frac{3\pi k}{4}\sinh\left((\frac{\pi}{\ga}-3)\frac{\pi k}{4})\right)\over
\sinh \pi k \sinh \left((\frac{\pi}{\ga}-2)\frac{\pi k}{4}\right)}
\Bigg\} \,. 
\ee 
Let us compare this result with the soliton reflection amplitude 
$P_{+}(\theta \,, \xi)$
of the boundary SG model with Dirichlet boundary conditions \cite{GZ}, 
which has the integral representation \cite{FS} 
\be
\frac{1}{i}\frac{d}{d\theta}\ln P_{+}(\theta\,, \xi)
=\int_{-\infty}^{\infty}dk\, e^{-i k \theta} \left[
{\sinh\left((1+ \frac{2\xi}{\pi \lambda})\frac{\pi k}{2}\right)\over
2 \cosh\frac{\pi k}{2} \sinh \frac{\pi k}{2\lambda}} +
{\sinh \frac{3\pi k}{4}\sinh\left((\frac{1}{\lambda}-1)\frac{\pi k}{4})\right)\over
\sinh \pi k \sinh \frac{\pi k}{4\lambda}} \right]
\,,
\ee 
where $\lambda$ and $\xi$ are the bulk and boundary IR parameters,
respectively.  Recalling the bulk ``lattice $\leftrightarrow$ IR''
relation (\ref{bulklatIR}), and assuming the boundary ``lattice
$\leftrightarrow$ IR'' relation
\be
\eta_{\pm} = \frac{1}{2}(\pi+ \ga) + 
\left(1-\frac{2\ga}{\pi}\right)\xi_{\pm} \,,
\label{boundlatIR}
\ee 
we conclude that 
\be
e^{i\Pf_{bdry}(\theta_{h}) +i \chi(2\theta_{h})} = P_{+}(\theta_{h} \,, 
\xi_{-})\, P_{+}(\theta_{h} \,, \xi_{+}) \,. \label{SGpart}
\ee 

Combining the results (\ref{prodRR}),  (\ref{calKresult})  
and (\ref{SGpart}), we conclude that the NLIE generates the
following SSG soliton boundary $S$ matrices
\be
R(\theta_{h}\,; \lambda\,, \xi_{\pm}) \sim P_{+}(\theta_{h} \,, 
\xi_{\pm}) \, P_{min}(\theta_{h}) \,. \label{IRresult}
\ee 
This is the boundary $S$ matrix which was proposed 
by Bajnok {\it et al.} \cite{BPT} for the the Dirichlet
BSSG${}^{+}$ model.
This is another of our main results. Similarly to the bulk case, 
the SSG boundary $S$ matrix is a product of SG and RSOS
boundary $S$ matrices. 

For general values of boundary parameters, the BSSG${}^{+}$ model with
one boundary has the conserved supercharge \cite{Ne, AK, BPT}
\be
\tilde Q_{+} = Q+ \bar Q + \ga' \Gamma \,,
\label{supercharge}
\ee
where $\Gamma=(-1)^{F}$ is the Fermionic parity operator, and $\ga'$
is an undetermined parameter.  \footnote{This parameter is called $\ga$ in
\cite{BPT}; however, here we add a prime in order to distinguish it
from our bulk lattice parameter.} Bajnok {\it et al.} also propose 
the relation
\be
\tilde Q_{+}^{2}=2\left(\tilde H + m \tilde Z \right) \,,
\label{Qsquared}
\ee 
where $\tilde H$ is the Hamiltonian,
and $\tilde Z$ is the topological charge.  Since the ground state has
$\tilde Q_{+} = \ga'$ and $\tilde Z = 0$,
it has energy $\tilde H  = \ga'^{2}/2$.  Our result 
(\ref{SSGboundvacuumenergy}) that each boundary contributes
vacuum energy $m/2$ implies that 
\be
\ga'= \pm \sqrt{m} \,, \label{gaprimeresult}
\ee
at least for the Dirichlet case. That is, we have succeeded to fix 
the undetermined parameter in the scattering theory proposed in \cite{BPT}.


\section{Ultraviolet limit}\label{sec:uvlimit}

In this Section we first analytically compute the Casimir energy
$E_{C}(L)$ (\ref{Casimirspin1}) in the UV limit $m L \rightarrow 0$.
The result (\ref{ECresult}) is proportional to the effective central charge 
\be
E_{C}(0)=-\frac{\pi}{24L}c_{eff}(0)   \,,
\label{ceffEC}
\ee 
where $c_{eff}(0) = c - 24 \Delta_{0}$, $c$ is the central charge, and
$\Delta_{0}$ is the $L_{0}$ eigenvalue of the ground state.
We then compare this result for $c_{eff}(0)$ to the value for the conformal limit of 
the SSG model with Dirichlet boundary conditions. In this way, we obtain 
a boundary ``UV $\leftrightarrow$ lattice'' relation (\ref{boundUVlat}).
When combined with the boundary ``lattice $\leftrightarrow$ IR''
relation from the previous section (\ref{boundlatIR}), we obtain the 
boundary ``UV $\leftrightarrow$ IR'' relation (\ref{boundUVIR}).

\subsection{NLIE computation}\label{subsec:nliecasimir}

We now proceed to analytically evaluate the Casimir energy
in the UV limit $m L \rightarrow 0$.
As is well known, only large values of
$|\theta|$ contribute in this limit. Let us first consider $\theta 
\gg 1$, and define the finite rapidity $\hat \theta$ by
\be
\theta = \hat \theta - \ln ( m L) \,.
\ee
The corresponding contribution $E_{C}^{+}$ to the Casimir energy
(\ref{Casimirspin1}) is given by
\be
L\, E_{C}^{+} = \frac{1}{16\pi}\int_{-\infty}^{\infty}d\hat \theta\, 
2i e^{\hat \theta}\left[ \ln \Bf_{+}(\hat \theta) - 
\ln \bar \Bf_{+}(\hat \theta) \right] \,,
\label{ECplus}
\ee
where the auxiliary functions $\Bf_{+}(\hat \theta) \equiv \Bf(\hat 
\theta - \ln ( m L))$, etc. satisfy the NLIE equations  
\be
\ln \bff_{+}(\hat \theta) &=& 
\int_{-\infty}^{\infty}d\hat\theta'\ \Gf(\hat\theta-\hat\theta'-i\varepsilon) 
\ln \Bf_{+}(\hat\theta'+i\varepsilon)  
-\int_{-\infty}^{\infty}d\hat\theta'\ \Gf(\hat\theta-\hat\theta'+i\varepsilon) 
\ln \bar \Bf_{+}(\hat\theta'-i\varepsilon)\non \\
&+& \int_{-\infty}^{\infty}d\hat\theta'\ \Gf_{2}(\hat\theta-\hat\theta'+i\varepsilon) 
\ln \Yf_{+}(\hat\theta'-i\varepsilon) + i e^{\hat\theta} + 
i\Pf_{bdry}(\infty) -i \pi\,, 
\non \\
\ln \bar \bff_{+}(\hat\theta) &=& 
-\int_{-\infty}^{\infty}d\hat\theta'\ \Gf(\hat\theta-\hat\theta'-i\varepsilon) 
\ln \Bf_{+}(\hat\theta'+i\varepsilon) 
+\int_{-\infty}^{\infty}d\hat\theta'\ \Gf(\hat\theta-\hat\theta'+i\varepsilon) 
\ln \bar \Bf_{+}(\hat\theta'-i\varepsilon)\non \\
&+& \int_{-\infty}^{\infty}d\hat\theta'\ 
\Gf_{2}(\hat\theta'-\hat\theta+i\varepsilon) 
\ln \Yf_{+}(\hat\theta'+i\varepsilon) - i e^{\hat\theta} - 
i\Pf_{bdry}(\infty) +i \pi
\,, \label{NLIEplus} \\
\ln \yf_{+}(\hat \theta) &=& \int_{-\infty}^{\infty}d\hat \theta'\ 
\Gf_{2}(\hat \theta-\hat \theta'+i\varepsilon) 
\ln \bar \Bf(\hat \theta'-i\varepsilon) 
+ \int_{-\infty}^{\infty}d\hat \theta'\ 
\Gf_{2}(\hat \theta'-\hat \theta+i\varepsilon) 
\ln \Bf(\hat \theta'+i\varepsilon) \non  \,,
\ee
which are independent of $m L$.  The $\Pf_{y}$ term (\ref{Pytheta}) 
does not appear in the third equation due to the fact 
$\Pf_{y}(\infty)=0 \ \mbox{ mod } 2\pi$.
We note here for later reference that 
\be
\Pf_{bdry}(\infty) = \frac{1}{2} \hat R(0) = \frac{\pi}{\pi-2\ga}
\left(\eta_{+}+\eta_{-}-3\ga\right) \,.
\label{tilderifnty}
\ee 

We use the so-called dilogarithm trick \cite{KP, Su1}. We  first rewrite the 
NLIE equations (\ref{NLIEplus}) in matrix form,
\be
u = v + K * w  \,,
\label{NLIEmatrix}
\ee
where
\be
u= \left( \begin{array}{c}
\ln \bff_{+}\\
\ln \bar \bff_{+}\\
\ln \yf_{+}
\end{array} \right) \,, \quad v= \left( \begin{array}{c}
                       i e^{\hat\theta} + i\Pf_{bdry}(\infty) - i \pi\\
                      -i e^{\hat\theta} - i\Pf_{bdry}(\infty) + i \pi\\
                        0
                  \end{array} \right) \,, \quad w=\left( \begin{array}{c}
                             \ln \Bf_{+}\\
			     \ln \bar \Bf_{+}\\
			     \ln \Yf_{+}
			     \end{array}\right) \,,
\ee 
the kernel $K$ is symmetric (i.e., 
$K_{ij}(\hat\theta\,,\hat\theta')=K_{ji}(\hat\theta'\,,\hat\theta)$), 
and the star $*$ denotes convolution. 
We see from (\ref{NLIEmatrix}) that 
\footnote{Here the prime denotes differentiation with respect to 
$\hat\theta$.}
\be
\int_{-\infty}^{\infty}d\hat\theta \left(w^{T}u' -w'^{T}u \right)= 
 \int_{-\infty}^{\infty}d\hat\theta \left(w^{T}v'  - 
w'^{T}v \right) \,, \label{trick}
\ee 
since the symmetry of the kernel implies that
\be
\int_{-\infty}^{\infty}d\hat\theta \left( w^{T} K'*w  - w'^{T} K * 
w\right) = 0 \,.
\ee
It follows from (\ref{trick}) that
\be
\lefteqn{\int_{-\infty}^{\infty}d\hat\theta\, \Big\{ (\ln \bff_{+})' \ln \Bf_{+} - \ln 
\bff_{+} (\ln \Bf_{+})' + (\ln \bar \bff_{+})' \ln \bar \Bf_{+} - \ln 
\bar \bff_{+} (\ln \bar \Bf_{+})'} \non \\
& & + (\ln \yf_{+})' \ln \Yf_{+} - \ln 
\yf_{+} (\ln \Yf_{+})' \Big\}
= \int_{-\infty}^{\infty}d\hat\theta\,  2i e^{\hat\theta} \left( \ln \Bf_{+} 
- \ln \bar \Bf_{+} \right) \non \\
& &-i \left(e^{\hat\theta}+\Pf_{bdry}(\infty) - \pi
\right)\left( \ln \Bf_{+} - \ln \bar \Bf_{+} \right) 
\Big\vert^{\hat\theta = \infty}_{\hat\theta = -\infty} \,. 
\label{trick2}
\ee
Since $\Bf_{+}=1+\bff_{+}$, etc., the LHS of (\ref{trick2}) can be
expressed in terms of the dilogarithm function $L_{+}(x)$, defined by
\be
L_{+}(x) = \frac{1}{2}\int_{0}^{x}dy \left[\frac{1}{y}\ln (1+y) - 
\frac{1}{1+y}\ln y \right] \,.
\ee 
The integral on the RHS of (\ref{trick2}) is essentially the 
sought-after quantity $L E_{C}^{+}$ (\ref{ECplus}). We conclude that
\be
L\, E_{C}^{+}&=& \frac{1}{8\pi}\Big\{ L_{+}\left(\bff_{+}(\infty)\right) - 
L_{+}\left(\bff_{+}(-\infty)\right) + L_{+}\left(\bar 
\bff_{+}(\infty)\right) - L_{+}\left(\bar \bff_{+}(-\infty)\right)  \label{ECplus2} \\
&+& L_{+}\left(\yf_{+}(\infty)\right) - L_{+}\left(\yf_{+}(-\infty)\right) 
+\frac{i}{2} \left(e^{\hat\theta}+\Pf_{bdry}(\infty) -\pi
\right)\left( \ln \Bf_{+} - \ln \bar \Bf_{+} \right) 
\Big\vert^{\hat\theta = \infty}_{\hat\theta = -\infty}
\Big\} \,. \non
\ee 

The plateau values of the auxiliary functions can be obtained from the
NLIE equations (\ref{NLIEplus}).  For $\hat\theta \rightarrow \infty$,
we readily obtain
\be
\bff_{+}(\infty) = \bar \bff_{+}(\infty) = 0 \,, \qquad \yf_{+}(\infty) = 1 \,.
\ee
To determine the plateau values for $\hat\theta \rightarrow -\infty$ 
is less trivial, as the corresponding plateau equations are nonlinear. We make the Ansatz
\be
\bff_{+}(-\infty) = e^{i \omega} \left(1+ e^{i \omega} \right)  \,, \quad
\bar \bff_{+}(-\infty) = e^{-i \omega} \left(1+ e^{-i \omega} \right) \,, 
\quad \yf_{+}(-\infty) = 1 + e^{i \omega}+ e^{-i \omega} \,,
\label{ansatz}
\ee 
where $\omega$ is still to be determined. We indeed find a solution 
with
\be
\omega = \frac{\Pf_{bdry}(\infty) - \pi}{2\left[\frac{3}{4}-\widehat {G}(0) 
\right]} = 2\left(\eta_{+}+\eta_{-}-\ga-\pi\right)  \,, \label{omega}
\ee
where we have used the result (\ref{tilderifnty}).
For the plateau values (\ref{ansatz}), the following sum 
rule holds \cite{Su2}
\be
L_{+}\left(\bff_{+}(-\infty)\right) + L_{+}\left(\bar \bff_{+}(-\infty)\right)  
+ L_{+}\left(\yf_{+}(-\infty)\right) - L_{+}(1) = \frac{\pi^{2}}{4} 
\,, \quad  |\omega| < \frac{2\pi}{3} \,.
\ee
The above bound is satisfied when the boundary parameters are in the domain 
(\ref{bpdomainspin1}).
Noting also that $L_{+}(0)=0$, we conclude that the Casimir energy is given by
\be
L\, E_{C}(0) = 2\,L E_{C}^{+} = -\frac{\pi}{24} \left[ \frac{3}{2} - 
\frac{12}{\pi (\pi-2\ga)} 
\left(\eta_{+}+\eta_{-}-\ga-\pi\right)^{2} \right] \,.
\label{ECresult}
\ee 
In obtaining the first equality, we have used the fact that the
contribution $E_{C}^{-}$ from $\theta \ll -1$ is the same as
$E_{C}^{+}$, and that $E_{C}(0)=E_{C}^{+}+E_{C}^{-}$.

\subsection{CFT analysis and UV-IR relation}\label{subsec:CFT}

Our result for the Casimir energy (\ref{ECresult}) together with the 
relation (\ref{ceffEC}) evidently imply that the effective central 
charge has the value
\be
c_{eff}(0) = c - 24 \Delta_{0} = \frac{3}{2} - 
\frac{12}{\pi (\pi-2\ga)} 
\left(\eta_{+}+\eta_{-}-\ga-\pi\right)^{2} \,.
\label{ceffresult}
\ee
We remark that for boundary parameter values
$\eta_{+}=\eta_{-}=\frac{1}{2}(\pi+\ga)$, the boundary terms 
in the Hamiltonian (\ref{diagbt}) which are
proportional to $S^{z}$ vanish; and also $\Delta_{0}$
vanishes, so that $c_{eff}(0) = c = 3/2$.  A similar phenomenon was
observed for the spin-1/2 case in \cite{LMSS}.

In the UV limit, the boundary SSG model with Dirichlet boundary
conditions (\ref{ssg}), (\ref{Dirichlet}) evidently reduces to a
system of one free Boson and one free Majorana Fermion, each with
Dirichlet boundary conditions.  For the former model, the central
charge and lowest dimension are given by \cite{Sa}
\be
c_{B}=1 \,, \qquad \Delta_{B}=\frac{1}{2\pi}\left(\varphi_{-} - 
\varphi_{+}\right)^{2} \,, 
\ee
while for the latter model \footnote{Since both the left and right 
boundaries have the same (Dirichlet) boundary condition, the operator content 
includes the identity operator, which has dimension zero.}
\be
c_{F}=\frac{1}{2} \,, \qquad \Delta_{F}=0 \,.
\ee
It follows that the SSG model with Dirichlet boundary conditions should have
\be
c=c_{B}+c_{F}=\frac{3}{2} \,, \qquad 
\Delta_{0}=\Delta_{B}+\Delta_{F}=\frac{1}{2\pi}\left(\varphi_{-} - 
\varphi_{+}\right)^{2} \,.
\ee
Comparing this CFT result with the NLIE result (\ref{ceffresult})
and recalling the {\it bulk} ``UV $\leftrightarrow$ lattice'' relation
(\ref{bulkUVlat}), we obtain the {\it boundary} ``UV $\leftrightarrow$
lattice'' relation
\be
\eta_{\pm} = \frac{1}{2}(\pi+\ga) \pm \frac{\beta}{2} \varphi_{\pm} \,.
\label{boundUVlat}
\ee 
Combining this result with the boundary ``lattice $\leftrightarrow$ IR''
relation (\ref{boundlatIR}), we finally arrive at the SSG boundary 
``UV $\leftrightarrow$ IR'' relation
\be
\xi_{\pm} = \pm \frac{2\pi}{\beta}\varphi_{\pm} \,.
\label{boundUVIR}
\ee 
This is another of our main results.

The relation (\ref{boundUVIR}) is similar to the one found by Ghoshal and 
Zamolodchikov \cite{GZ} for the SG model (namely, $\xi = 
\frac{4\pi}{\beta}\varphi_{0}$), and it can be understood in a similar way.
Indeed, for the SSG model, it is also plausible to assume a linear
relation between these parameters, 
\be
\xi_{\pm}= a + b\varphi_{\pm} \,.
\ee 
When $\varphi_{\pm}=0$, the model has the symmetry $\varphi \mapsto
-\varphi$; and, since the RSOS factor of the boundary $S$ matrix is
proportional to the identity for BSSG$^{+}$ \cite{BPT}, the soliton and
antisoliton reflection amplitudes should be equal, which corresponds
to $\xi_{\pm}=0$.  Thus, $a=0$.  Furthermore, as in the SG model,
there are boundary bound states corresponding to poles of the 
boundary $S$ matrix at $\theta = i \nu_{n}$, where
\be
\nu_{n} = \frac{\xi_{\pm}}{\lambda} -\frac{(2n+1)\pi}{2\lambda} \,, 
\quad n =0 \,, 1\,, \ldots .
\label{poles}
\ee 
Moreover, these states satisfy \cite{BPT} 
\be
\tilde Q_{+}^{2}=2\left(\frac{\ga'^{2}}{2} + m \cos \nu_{n} + m \right) \,,
\label{Qsquaredboundstate}
\ee 
which is consistent with (\ref{Qsquared}), since these states have 
$\tilde Z=1$.  When $\varphi_{\pm}$ have the half-period
values $\varphi_{\pm}=\pm \pi/\beta$ \footnote{The Lagrangian
(\ref{ssg}) has the periodicity $\varphi \mapsto \varphi +
2\pi/\beta$.}, 
the $n=0$ bound state should have the same $\tilde
Q_{+}^{2}$ eigenvalue as the ground state.  From
(\ref{Qsquaredboundstate}), we see that this condition corresponds to
$\nu_{0}=\pi$, which in turn implies
$\xi_{\pm}=\pi(\lambda+\frac{1}{2}) = 2\pi^{2}/\beta^{2}$, as follows
from (\ref{poles}) and the bulk UV-IR relation (\ref{bulkUVIR}).  It
follows that $b= \pm 2\pi/\beta$, and so we recover the boundary UV-IR
relation (\ref{boundUVIR}).

\section{Intermediate volume and BCPT}\label{sec:intermediate}

\begin{figure}[htb]
    \centering
	    \includegraphics[width=0.60\textwidth]{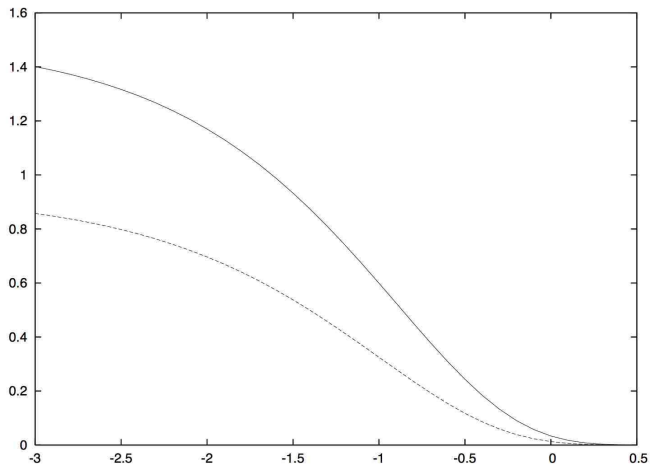}
	    \caption[xxx]{\parbox[t]{0.6\textwidth}{
	    $c_{eff}$ vs. $\ln l$, for $\ga=2\pi/7$,
	    $\eta_{+}=\eta_{-}= (\pi+\ga)/2$ (solid line)
	    and $\eta_{+}=\eta_{-}= \pi/2+ 3\ga/4$ (dotted line)}
	}
	\label{fig:ceffvslogl}
\end{figure}

For intermediate values of volume $l \equiv m L$, the Casimir energy
cannot be computed analytically.  Nevertheless, it is possible to
solve the NLIEs (\ref{NLIEspin1coord}) numerically by iteration, and
evaluate the Casimir energy through (\ref{Casimirspin1}).  Two sample
plots of $c_{eff}(l) \equiv -24 L E_{C}(L)/\pi$ vs.  $\ln l$ are shown in Figure
\ref{fig:ceffvslogl}.  The numerical result for $c_{eff}(l)$ in the UV region $l
\rightarrow 0$ coincides with the analytical result
(\ref{ceffresult}).  Also, as expected, $c_{eff}(l)$ decreases
monotonically to 0 as $l$ varies from the UV region to the IR region
$l \rightarrow \infty$.

For small values of $l$, these numerical results can be compared with
those from boundary conformal perturbation theory (BCPT).  We shall
follow closely the presentation in the Appendix of the second
reference in \cite{AN}. We regard the SSG
model (\ref{ssg}) as a perturbed boundary CFT,
\be
L = L_{BCFT} + L_{pertb} \,, \qquad 
L_{pertb} = -\frac{m_{0}}{2}\int_{0}^{L}dx\, \Phi(x,t) \,,
\ee
where $L_{BCFT}$ is the Lagrangian for a free scalar field $\varphi$
and a free Majorana Fermion field $\Psi$ obeying Dirichlet boundary
conditions (\ref{Dirichlet}).
We restrict our attention here to the particular
case $\varphi_{-}= \varphi_{+} \equiv \varphi_{0}$, in order to avoid 
introducing boundary-changing operators. The bulk perturbing operator 
$\Phi(x,t) = \cos(\beta \varphi) \bar \Psi \Psi$ is a primary field with weights
$(\Delta, \Delta)$, where
\be 
\Delta  = \frac{1}{2} + \frac{\beta^{2}}{8\pi} \,, \qquad 
\frac{1}{2} < \Delta < 1 \,.
\ee
The corresponding Hamiltonian is
\be
H(L) = H_{BCFT} + \frac{m_{0}}{2}\int_{0}^{L}dx\, \Phi(x,t) \,.
\ee
We map the infinitely long strip of width $L$ to the upper half
plane by setting $z=e^{i\frac{\pi}{L}(x+t)}$, where $t=-iy$ is the
Euclidean time.  Taking the Hamiltonian at $t=0$ and changing the
integration variable to $\theta=\frac{\pi}{L}x$, we have
\be
H(L) = \frac{\pi}{L}\left(L_{0}-\frac{c}{24}\right) + 
\frac{m_{0}}{2}\left(\frac{\pi}{L}\right)^{2\Delta-1}\int_{0}^{\pi}
d\theta\, \Phi(e^{i\theta}, e^{-i\theta}) \,,
\ee
with $c=3/2$.
First-order perturbation theory implies that the ground-state energy 
is given by
\be
E_{0}(L) = \frac{\pi}{L}\left(\Delta_{0}-\frac{c}{24}\right) + 
\frac{m_{0}}{2}\left(\frac{\pi}{L}\right)^{2\Delta-1}\int_{0}^{\pi}
d\theta\, \langle 0| \Phi(e^{i\theta}, e^{-i\theta}) |0 \rangle + 
O(m_{0}^{2}) \,,
\ee
where $|0\rangle$ is the ground state of the unperturbed theory. 
Using the fact that the bulk one-point function has the form 
\be
\langle 0| \Phi(e^{i\theta}, e^{-i\theta}) |0 \rangle 
=\frac{c_{bulk}}{(2\sin \theta)^{2\Delta}} \,,
\ee
and performing the $\theta$ integration, we obtain
\be
E_{0}(L) = -\frac{\pi}{24L}c_{eff}(0) + 
\frac{m_{0}}{4}\left(\frac{\pi}{2L}\right)^{2\Delta-1}
\frac{\Gamma(\frac{1}{2}-\Delta) \Gamma(\frac{1}{2})}{\Gamma(1-\Delta)}c_{bulk}
+ O(m_{0}^{2}) \,,
\ee
where $c_{eff}(0)=c -24 \Delta_{0}$, as before. The energy
$E_{0}(L)$ is the sum of bulk, boundary and Casimir energies
\be
E_{0}(L) = E_{B} L + E_{b} + E_{C}(L) \,, \qquad
E_{C}(L) = -\frac{\pi}{24L}c_{eff}(l)
\,.
\ee
Recalling the bulk and boundary energy results (\ref{SSGbulkvacuumenergy}), 
(\ref{SSGboundvacuumenergy}), we obtain
\be
c_{eff}(l) = c_{eff}(0) + \frac{24 l}{\pi}  
- 3m_{0}\left(\frac{\pi}{2L}\right)^{2\Delta-2}
\frac{\Gamma(\frac{1}{2}-\Delta) \Gamma(\frac{1}{2})}{\Gamma(1-\Delta)}c_{bulk}
+ O(m_{0}^{2}) \,. \label{e0nlie1}
\ee 

Following Cardy and Lewellen \cite{CL}, the coefficient of the bulk
one-point function can be expressed as a ratio of scalar products,
\be
c_{bulk} =\frac{\langle \!\langle \Phi| D \rangle}{\langle \!\langle 0| D \rangle}
= \frac{\langle \!\langle \cos (\beta \varphi) | 
B_{D}(\varphi_{0}) \rangle}
{\langle 0, 0| B_{D}(\varphi_{0}) \rangle}\Big\vert_{c=1}
\cdot
\frac{\langle \!\langle \bar\Psi \Psi | \tilde 0 \rangle}
{\langle \!\langle 0| \tilde 0 \rangle}\Big\vert_{\rm Ising} \,.
\label{cbulk1}
\ee 
Here $| B_{D}(\varphi_{0}) \rangle$ is the $c=1$ Dirichlet boundary state 
\cite{Sa}
\be
| B_{D}(\varphi_{0}) \rangle = {\cal N}_{D}
\sum_{k=-\infty}^{\infty}e^{i k \beta 
\varphi_{0}}e^{-\sum_{n=1}^{\infty}\frac{1}{n}\alpha_{-n} \bar 
\alpha_{-n}} |0, k \rangle \,,
\ee
where $|0, k \rangle$ is annihilated by $\alpha_{n}$ and $\bar 
\alpha_{n}$ for $n>0$, and has weight $h_{0,k}=\frac {k^{2} 
\beta^{2}}{8\pi}$; and ${\cal N}_{D}$ is a normalization factor.
Hence, $e^{\pm \beta \varphi}$ (which have weight $\frac 
{\beta^{2}}{8\pi}$) are identified with $|0, \pm 1\rangle$.
Thus, $|\cos (\beta \varphi)\rangle \!\rangle= 
\frac{1}{2}\left( |0, 1\rangle + |0, -1\rangle \right)$, and we 
conclude that
\be
\frac{\langle \!\langle \cos (\beta \varphi) | 
B_{D}(\varphi_{0}) \rangle}
{\langle 0, 0| B_{D}(\varphi_{0}) \rangle}\Big\vert_{c=1} 
= \cos (\beta \varphi_{0}) \,.
\label{cbulk2}
\ee 
Moreover, $\bar \Psi \Psi = -\frac{1}{\pi}\varepsilon$, where $\varepsilon$ 
is the Ising energy density operator.
Recalling the expressions  \cite{Ca} for the boundary states corresponding to 
fixed boundary conditions in terms of Ishibashi states
\be
| \tilde 0 \rangle &=& \frac{1}{\sqrt 2}| 0 \rangle \!\rangle + 
\frac{1}{\sqrt 2}| \varepsilon \rangle \!\rangle
+ \frac{1}{\sqrt[4] 2}| \sigma \rangle \!\rangle \,, \non \\
| \tilde {\frac{1}{2}} \rangle&=& \frac{1}{\sqrt 2}| 0 \rangle 
\!\rangle + 
\frac{1}{\sqrt 2}| \varepsilon \rangle \!\rangle
- \frac{1}{\sqrt[4] 2}| \sigma \rangle \!\rangle \,,
\ee 
we obtain \footnote{We evidently obtain the same result if we consider 
$| \tilde {\frac{1}{2}} \rangle$ instead of $| \tilde 0 \rangle$.}
\be
\frac{\langle \!\langle \varepsilon | \tilde 0 \rangle}
{\langle \!\langle 0| \tilde 0 \rangle}\Big\vert_{\rm Ising} = 1\,.
\label{cbulk3}
\ee 
We conclude from (\ref{cbulk1}), (\ref{cbulk2}) and (\ref{cbulk3})
that the coefficient of the bulk one-point function is given by
\be
c_{bulk} = -\frac{1}{\pi}\cos (\beta \varphi_{0}) \,. \label{cbulkfinal}
\ee
Moreover, the parameter $m_{0}$ is related to the SSG soliton mass $m$ by the 
so-called mass-gap formula \cite{BF}
\be
m_{0}=\frac{8\Gamma(\Delta)}{\Gamma(1-\Delta)}
\left[\frac{\pi m}{8}\left(\frac{2\Delta-1}{1-\Delta}\right)\right]^{2-2\Delta}\,.
\label{massgap}
\ee
This relation has the correct classical limit $\beta \rightarrow 0$, 
namely $m_{0} \rightarrow m_{1}$, where $m_{1}$ is the mass of the 
first breather (\ref{breathermass}).

Substituting (\ref{cbulkfinal}) and (\ref{massgap}) into  
(\ref{e0nlie1}), we obtain the result 
\be
c_{eff}(l) \approx c_{eff}(0) +  \frac{24}{\pi} \left(
l - \alpha_{1}\, l^{2-2\Delta} + 
\alpha_{2}\, l^{4-4\Delta} \right) \,, \qquad l \ll 1 \,, 
\label{BCPTresult1}
\ee 
where
\be
\alpha_{1} = -\left[\frac{2\Delta-1}{4(1-\Delta)}\right]^{2-2\Delta}
\frac{\Gamma(\frac{1}{2}-\Delta) \Gamma(\frac{1}{2}) 
\Gamma(\Delta)}{\Gamma(1-\Delta)^{2}}
\cos (\beta \varphi_{0}) \,.
\label{BCPTresult2}
\ee 
We have not attempted to compute the second-order correction 
$\alpha_{2}$. 
In Table \ref{table:alpha1}, we compare the analytical BCPT result for
$\alpha_{1}$ (\ref{BCPTresult2}) with results obtained by fitting
numerical NLIE values for $c_{eff}(l)$ to the curve
(\ref{BCPTresult1}).  The excellent agreement between the analytical
and numerical values further supports the validity of our NLIE
(\ref{NLIEspin1coord}), as well as the boundary energy result
(\ref{SSGboundvacuumenergy}) and the boundary ``UV $\leftrightarrow$
lattice'' relation (\ref{boundUVlat}).

\begin{table}[htb] 
  \centering
  \begin{tabular}{|c|c|c|c|c|c|c|}\hline
    &\multicolumn{2}{c|}{$\varphi_{0}=0$}  &\multicolumn{2}{c|} 
    {$\varphi_{0}=0.1$} &\multicolumn{2}{c|} {$\varphi_{0}=0.2$}\\ \cline{2-7} 
    \rb{$\pi/\ga$} & NLIE & BCPT  & NLIE & BCPT  & NLIE & BCPT\\
    \hline
       2.6       &       1.05337       &     1.05314  & 1.03813 & 
       1.0379 & 0.992871 & 0.992644 \\
       2.8     &       0.977854      &     0.977706  & 0.960355 & 
       0.960207 & 0.908481 & 0.908335  \\
       3      &       0.898947       &     0.898868  & 0.880187 &  
       0.880108 & 0.824691 & 0.82461 \\
       3.5     &       0.717426      &     0.717416  & 0.698197 & 
       0.698184 & 0.641543 & 0.641519\\
       4       &   0.572666         &   0.572698     & 0.554784 &   
       0.5548 & 0.502248 & 0.502225\\
       4.5  &  0.46177 & 0.461973  & 0.445794 & 0.445941 & 0.398955 & 0.398957 \\
       5 & 0.37713  &  0.377594 & 0.363094 & 0.363448 & 0.321985 & 0.322071 \\
         \hline
        \end{tabular}
        \caption[xxx]{\parbox[t]{0.8\textwidth}{
	Comparison of NLIE (numerical) and BCPT (analytical) results for $\alpha_{1}$, 
        for various values of bulk and boundary parameters.}
	}
       \label{table:alpha1}
     \end{table}

\section{Discussion}\label{sec:conclude}

We have proposed a set of nonlinear integral equations
(\ref{NLIEspin1coord}) to describe the boundary supersymmetric sine-Gordon
model BSSG${}^{+}$ with Dirichlet boundary conditions on a finite
interval (\ref{ssg}), (\ref{Dirichlet}).  In particular, we have found
the boundary terms $\Pf_{bdry}(\theta)$ and $\Pf_{y}(\theta)$
which encode boundary effects.  We have computed the corresponding boundary $S$
matrix (\ref{IRresult}), and found that it coincides with the one
proposed by Bajnok {\it et al.} \cite{BPT} for the Dirichlet
BSSG${}^{+}$ model, with conserved supercharge (\ref{supercharge})
depending on a parameter $\ga'$. We have determined this parameter 
(\ref{gaprimeresult}) by computing the boundary vacuum energy. We have also
proposed a relation (\ref{boundUVIR}) between the (UV) parameters in
the boundary conditions and the (IR) parameters in the boundary $S$
matrix.  Moreover, we have demonstrated that the NLIEs can be solved
numerically for intermediate values of $mL$, and we have found
agreement with our analytical result (\ref{ceffresult}) for the
effective central charge in the UV limit and with boundary 
conformal perturbation theory (\ref{BCPTresult1}), (\ref{BCPTresult2}).

There are a number of related questions which remain to be addressed.
While we have focused here primarily on the ground state, it should be
interesting to study bulk (along the lines \cite{HRS}) and also
boundary excitations.  It would also be interesting to formulate the
TBA equations for this model, and compare the results with those from
our NLIE.

Ultimately, we would like to extend this (Dirichlet) analysis of
BSSG${}^{+}$ to the case of general integrable boundary conditions
\cite{Ne}.  For this we would need the Bethe Ansatz solution of the
open spin-1 XXZ quantum spin chain with general integrable boundary
terms \cite{IOZ2}, which is currently under investigation.

It is not evident which integrable spin-1 model, if any, corresponds
to the Dirichlet BSSG${}^{-}$ model.  Indeed, the boundary terms
(\ref{diagbt}) already correspond to the most general diagonal
$c$-number solution of the boundary Yang-Baxter equation.  Perhaps
additional dynamical boundary degrees of freedom or impurities will be
required.  An interesting feature of BSSG${}^{-}$ is that, in contrast
to BSSG${}^{+}$, the RSOS part of the boundary $S$ matrix \cite{BPT}
{\it does} depend on the boundary parameter.  Therefore, the
corresponding NLIE may have a $\Pf_{y}(\theta)$ term which
also depends on the boundary parameters.

Another interesting problem is to identify the specific feature of the
spin-1 models which leads, in the continuum limit, to supersymmetry.
It may be possible to construct 
an integrable open spin-1 chain whose continuum limit is integrable
but {\it not} supersymmetric, such as models considered in \cite{IOZ}.
We hope to be able to address some of these questions in the future.

\section*{Acknowledgments}
We began this project at the APCTP Focus Program ``Finite-size
technology in low dimensional quantum field theory (II)'' in Pohang,
South Korea during summer 2005. One of us (CA) thanks Shizuoka 
University, SLAC and University of Miami for support. 
We are grateful to Z. Bajnok for 
helpful correspondence.
This work was supported in part by KOSEF-M60501000025-05A0100-02500 
(CA), by the
National Science Foundation under Grants PHY-0244261 and PHY-0554821 
(RN), and by the Ministry of Education of Japan, a Grant-in-Aid for 
Scientific Research 17540354 (JS).

\begin{appendix}
    
\section{Range of boundary parameters}\label{sec:domain}

We derive here the restrictions on the ranges of the boundary
parameters given in the main text (\ref{bpdomainspin12}),
(\ref{bpdomainspin1}).

\subsection{Spin-$\frac{1}{2}$}\label{sec:domainspinhalf}

Our argument depends on two assumptions.
First, the imaginary part of $\ln \bar{a}(x)$ is monotonically 
increasing.
Second, there should be no holes on the real axis (except for a single hole 
at the origin).
The first assumption suffers from exceptions for some values of
$\gamma$  if  the system size is small. 
However, we regard this as a finite size effect and
expect our assumption to be valid in the thermodynamic limit.

\begin{figure}[bthp]
\centering
{  \includegraphics[width=5cm]{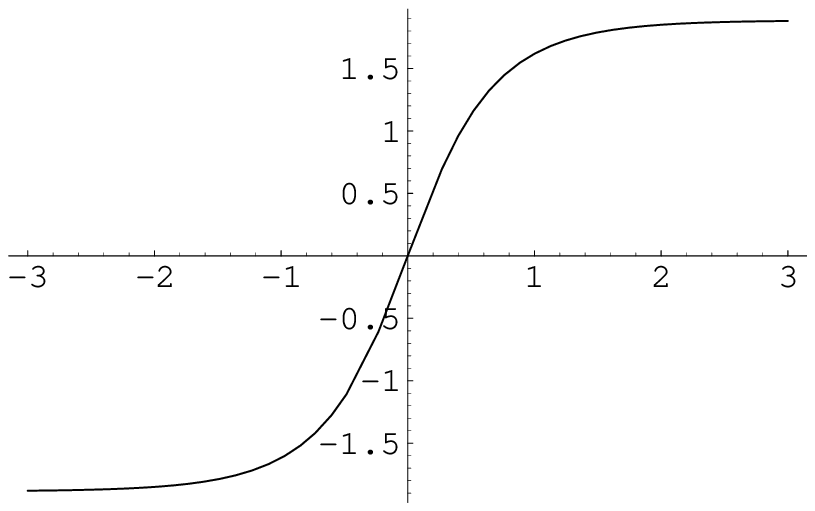} \hspace{1cm}
  \includegraphics[width=5cm]{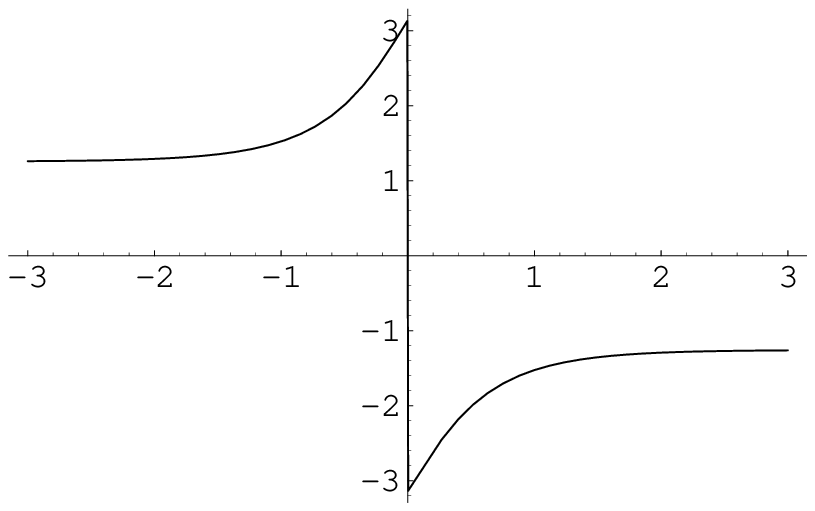} }
\caption{The functions $\vartheta_- $ (left) and $\vartheta_+$ (right).}
\label{phasepic}
\end{figure}

\begin{figure}[bthp]
\centering
 \includegraphics[width=5cm]{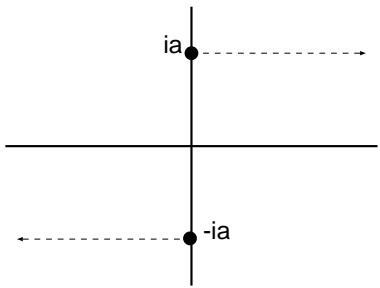} 
\caption{The branch cut lines for  $\vartheta_-$.}
\label{branchcutpic}
\end{figure}

We consider first the repulsive regime $0<\gamma<\frac{\pi}{2}$. 
Let  $0<a<\frac{\pi}{2}$.
We then define two functions,
$$
\vartheta_-(x, a) = \frac{1}{i}  \ln (-\frac{\sinh(x-i a )}{ 
\sinh(x+i a )}) \,, \qquad
\vartheta_+(x, a) =  \frac{1}{i}  \ln (\frac{\sinh(x-i a)}{ 
\sinh(x+i a)}) \,,
$$
where we adopt the convention for the logarithm such that its imaginary part is 
restricted to $[-\pi \,,\pi)$.
These functions are essentially the same but for the choice of phase
(Figure \ref{phasepic}).
We choose the branch of $\vartheta_-(x, a)$ such that a branch cut 
line emerges from 
$i a$ and goes to positive infinity, and another line starts from $-ia$ 
and goes to negative  infinity, as shown in Figure \ref{branchcutpic}.

\noindent 
In the rest of the range, $\frac{\pi}{2}<a<\pi$, we define
\begin{equation}
\vartheta_-(x, a) = -\vartheta_-(x, \pi-a) \,, \qquad
\vartheta_+(x, a) =  -\vartheta_+(x, \pi-a) \,.
\label{extendedtheta}
\end{equation}

Recalling (\ref{defa}),  we consider 
\begin{align}
\frac{1}{i} {\rm Log}\, \bar{a}(x)&= \sum_{\alpha=\pm} \vartheta_-(x, \frac{H_{\alpha} \gamma}{2}) 
+ \vartheta_+(2x,\gamma)
 + N  \sum_{\alpha=\pm} \vartheta_- (x- \alpha \Lambda, \frac{\gamma}{2})  \nonumber \\
 &-\sum_{k=1}^M \left[ \vartheta_-(x-v_k, \gamma)+  \vartheta_- 
 (x+v_k, \gamma) \right] \,,
\label{logabar}
\end{align}
where $M$ denotes the number of BAE roots.
By Log, we mean the logarithm  whose imaginary part is {\bf not} restricted to 
$[-\pi,\pi)$.
Strictly speaking,  $\bar{a}$ is defined in the upper half-plane.
We assume that it is also well-defined on the real axis.
Note that $\frac{1}{i} {\rm Log}\, \bar{a}(0^{+})=\vartheta_+(0^{+},\gamma)=
-\pi$ for $0<\gamma<\pi/2$.

With the help of the identities 
\be
\vartheta_{-}(\infty, a) = \pi - 2a \,, \qquad \vartheta_{+}(\infty, 
a) = - 2a \,, \qquad 0 < a < \frac{\pi}{2} \,,
\non
\ee 
one easily derives the asymptotic value
\begin{equation}
    \frac{1}{i\pi}{\rm Log}\, \bar{a}(\infty)=2(N-M+1) -\frac{\gamma}{\pi}(2N-4M +2+H_+ 
    + H_-) 
\label{lnabarinf}
\end{equation}
for $H_{\pm}$ in the in the interval (\ref{bpdomainspin120}).
We denote the integer part of the above as $n$,
$$
n =
2(N-M+1)-\delta \,,
$$
where
$$\delta= \lfloor  \frac{\gamma}{\pi }(2N-4M+2+H_+ + H_-) \rfloor +1 
\,,$$
and   $\lfloor x \rfloor$ specifies the integer
part of $x$.
When $x$ is a root or a hole, $\frac{1}{i\pi} {\rm Log}\, \bar{a}(x)$ is 
an odd integer.
The  range of  $\frac{1}{i\pi} {\rm Log}\, \bar{a}$
is set so as to be able to accommodate $M$ roots. 
As already noted, there should be no holes (except for a single hole 
at the origin).
Thanks to the assumption of monotonicity, these conditions
determine the range of $n$ ,
\begin{equation}
2M-1 \le n < 2M+1 \,.
\label{Mrestriction}
\end{equation}
This is equivalent to having $\delta=2,3$.
The ground state selects $M=N/2$, and this leads to the conclusion 
(\ref{bpdomainspin12}),
$$
\frac{\pi}{\gamma} < H_+ + H_- +2  < \frac{3\pi}{\gamma} \,.
$$

In the attractive regime $\pi/2<\gamma<\pi$, some functions in
(\ref{logabar}) must be rewritten in terms of
$\tilde\gamma=\pi-\gamma$ which is now in the range $(0,\pi/2)$.
Thanks to (\ref{extendedtheta}), this only results in an extra $2\pi$
on the RHS of (\ref {lnabarinf}).
  In this case, however,
$\frac{1}{i} {\rm Log}\,  \bar{a}(0^{+})=\pi $, which cancels this extra $2\pi$
in the argument for the possible values of $\delta$.
We then reach the same conclusion on the range of $H_{\pm}$.

\subsection{Spin-1}\label{sec:domainspin1}

The spin-1 case suffers from an extra complication 
due to the absence of an
auxiliary function which precisely encodes the ``branch cut integers''
such as $a$ and $\bar{a}$ in the spin-1/2 case.
Previous studies \cite{Su2, HRS} nevertheless give this interpretation to
$\ln b(x)$.  It roughly encodes the information of the two-string center.
We assume this here, and utilize $b$ in place of $\bar{a}$ in the above argument.

We consider first the repulsive regime, $0<\gamma<\frac{\pi}{3}$.
The following representation is convenient for our purpose (recall 
Eqs. (\ref{T1}), (\ref{bT1Q})),
\begin{align}
\Im m\, {\rm Log}\, b(x)&=
N\sum_{\alpha=\pm} \bigl[ \vartheta_-(x -\alpha \Lambda,\frac{\gamma}{2}) + 
\vartheta_-(x -\alpha \Lambda,\frac{3\gamma}{2}) \bigr] \nonumber  \\
&-\sum_{\alpha=\pm} \bigl[ \vartheta_-(x, \eta_{\alpha}-\frac{\gamma}{2}) + 
\vartheta_-(x, \eta_{\alpha}+\frac{\gamma}{2}) \bigr]
+\vartheta_+(2x,2 \gamma)  \nonumber  \\
&-\sum_{k=1}^M  \bigl[ \vartheta_-(x-v_k, \frac{\gamma}{2}) 
+\vartheta_-(x+v_k, \frac{\gamma}{2}) + \vartheta_-(x-v_k, \frac{3\gamma}{2}) 
+\vartheta_-(x+v_k, \frac{3\gamma}{2}) \bigr]  \nonumber \\
& +\Im m\,  \ln(1+ a_1(x-\frac{i\gamma}{2}))\,, 
\end{align}
where we define $a_1(x) = l_{1}(x)/l_{2}(x)$.
In the above, $\vartheta_+(2x,2 \gamma)$ should be understood as  $-\vartheta_+(2x,\pi-2 \gamma)$
if $\pi/4<\gamma<\pi/3$. 
The above expression is valid for $\Im m \, x =+ \epsilon$.

We assume that the ground state is given by a sea of  
2-strings with slight deviations, $|{\Im}m \, v_k|=\gamma/2 + 
\epsilon$, and $M=N$.
Then a  careful analysis concludes 
$$
\Im m\,  \ln (1+a_1 (x+i\epsilon - \frac{i\gamma}{2})) |_{x\rightarrow \infty}
= \gamma - (\eta_+  +\eta_-) +\delta \pi \,,
$$
where $\delta$ is an integer satisfying $ |\gamma - (\eta_+  +\eta_-) +\delta \pi|<\pi$.
This leads to the following expression of the asymptotic value,
$$
\Im m\, {\rm Log}\,  b(\infty) = 3 (\eta_+ + \eta_- -\gamma) - 4 \pi 
+ 2\pi N 
+\delta \pi \,,
$$
where, for simplicity, we have further restricted the boundary 
parameters to the interval $\pi/2 < \eta_{\pm} < \pi- \gamma/2$.
 
For the spin-1 case, we must locate
$N/2$  2-string centers on the real positive $x$-axis.
Then an argument similar to the one for the spin-1/2 case leads to
 $$
\pi < \eta_+ +\eta_- -\gamma +\frac{\delta}{3} \pi < \frac{5\pi}{3}.
 $$
 One can then easily draw a conclusion that  $\delta=1$ is the only 
 consistent choice,
 and that the constraint (\ref{bpdomainspin1}) must be imposed,
 $$
\frac{2\pi}{3} +\gamma < \eta_+ +\eta_ -< \frac{4\pi}{3}+\gamma \,.
 $$
A little modification leads to the same conclusion for the attractive 
regime $\pi/3<\gamma<\pi/2$.

\section{Boundary terms}\label{sec:boundterms}

We provide here details in the computation of various boundary terms 
appearing in the NLIEs.

\subsection{Spin-$\frac{1}{2}$:  $C_{T}(k)$}\label{subsec:details12}

In the spin-$\frac{1}{2}$ calculation, the quantity $C_{T}(k)$ is defined by (\ref{CTdef})
\be
C_{T}(k) = {\widehat G}(k)\, C(k)+D(k) \,.
\ee
Since $B^{(\pm)}(x)$ and $\phi(x\pm \frac{i\ga}{2})$ are ANZ near the real axis, the contour
integral on $C_1$ in (\ref{Ck}) can be changed into $-C_2$, and $C(k)$ becomes
\beqa
C(k)&=&-D(k)-\delta(k) \,, \label{CintermsofD} \\
\delta(k)&=&\oint_{C} \left[ \ln \mu(x) \right]'' e^{ikx} =-i k 
\oint_{C} \left[ \ln \mu(x) \right]' e^{ikx}
= 2 \pi k \,. \label{deltaresult}
\eeqa
It follows that
\beq
C_{T}(k)= D(k)\left[1-{\widehat G}(k) \right]- 2\pi k {\widehat G}(k) \,.
\label{CT1}
\eeq

We now evaluate $D(k)$, which is defined by (\ref{dkdef}), i.e.,
\bea
D(k)&=&\int_{C_2}dx\
\Bigg\{\ln \Bigg[{ \sinh(x+{i\ga H_{+}\over 2})\sinh(x+{i\ga H_{-}\over 2})
\sinh^{N}(x-\La+{i\ga\over 2}) \sinh^{N}(x+\La+{i\ga\over 2}) \over
\sinh(x-{i\ga H_{+}\over 2})\sinh(x-{i\ga H_{-}\over 2})
\sinh^{N}(x-\La-{i\ga\over 2}) \sinh^{N}(x+\La-{i\ga\over 
2})}\\
& & \qquad \times {\sinh(2x+i\ga)\over \sinh(2x-i\ga)}\Bigg]\Bigg\}'' e^{ikx} \,.
\eea
Making use of the identities (\ref{psi}) and (\ref{psi2}), we obtain
\be
D(k) &=&2\pi\psi(k)\Big\{
e^{\left({\ga H_{+}\over{2}}-\pi\right)k}+e^{\left({\ga H_{-}\over{2}}-\pi\right)k}
-e^{-{\ga H_{+}\over{2}}k}-e^{-{\ga H_{-}\over{2}}k} \non \\
&+&
N(e^{i\La k}+e^{-i\La k})\left[e^{\left({\ga\over{2}}-\pi\right)k}-e^{-{\ga\over{2}}k}
\right]
\Big\} 
+ 2\pi\psi_2(k)\left[e^{\left({\ga\over{2}}-{\pi\over{2}}\right)k}-e^{-{\ga\over{2}}k}
\right] \,. \label{Dk}
\ee
We have assumed here that the boundary parameters $H_{\pm}$ are in 
the domain (\ref{bpdomainspin12}).

Using these results we can simplify the first term in Eq. (\ref{CT1}). From (\ref{Gk}),
\beq
\psi(k)[1-{\widehat G}(k)]=-{k e^{\frac{\pi k}{2}}\over 
4\cosh({\ga k\over 2})\sinh\left((\ga-\pi)\frac{k}{2}\right)} \,, 
\quad
\psi_{2}(k)[1-{\widehat G}(k)]=-{k e^{\frac{\pi k}{4}}\cosh{\pi 
k\over 4}\over 
2\cosh({\ga k\over 2})\sinh\left((\ga-\pi)\frac{k}{2}\right)} \,. \non 
\eeq
Combining these,
\bea
{1\over{2\pi k}}D(k)[1-{\widehat G}(k)]&=&-{N\cos(\La k)
\over \cosh({\ga k\over 2})}
-{\cosh({\pi k\over 4}) \sinh\left((\ga-\frac{\pi}{2})\frac{k}{2}\right)\over
  \cosh({\ga k\over 2}) \sinh\left((\ga-\pi)\frac{k}{2}\right)} \\
&-&{\sinh\left((\ga H_{+}-\pi)\frac{k}{2}\right)
   +\sinh\left((\ga H_{-}-\pi)\frac{k}{2}\right)
\over 2\cosh({\ga k\over 2}) \sinh\left((\ga-\pi)\frac{k}{2}\right)} \,.
\eea

Finally, taking into account also the second term in (\ref{CT1}),
we conclude that $C_{T}(k)$ is given by
\be
C_{T}(k) &=& -2\pi k \Bigg\{ {N\cos(\La k)\over \cosh({\ga k\over 2})} 
+{\sinh\left((\ga H_{+}-\pi)\frac{k}{2}\right)
   +\sinh\left((\ga H_{-}-\pi)\frac{k}{2}\right)
\over 2\cosh({\ga k\over 2}) \sinh\left((\ga-\pi)\frac{k}{2}\right)} 
\non \\
& & +{\cosh({\ga k\over 4}) \sinh\left((2\ga-\pi)\frac{k}{4}\right)\over
  \cosh({\ga k\over 2}) \sinh\left((\ga-\pi)\frac{k}{4}\right)} 
\Bigg\} \,. \label{CTfinal}
\ee

\subsection{Spin-1: $C(k)$}\label{subsec:details1}

In the spin-1 calculation, the quantity $C(k)$ is defined by (\ref{spin1Cdef})
\be
C(k)=-\widehat G(k)\, D_q(k)+ 
\widehat G_2(k)\, \widehat{Lf''}(k)  +\widehat{LC''_b}(k)
- e^{-{\ga k\over 2}}\widehat{L\mu''}(k)
+{\delta(k)\over e^{\ga k\over 2}+e^{-{\ga k\over 2}}} \,.
\label{spin1Cdefagain}
\ee 
The quantity $D_q(k)$ (\ref{Dqdef}) can be evaluated as follows: 
\be
\widehat{Lt''_+}(k)&=&2\pi\psi_2(k)e^{-\ga k}
+\left(e^{{\ga k\over{2}}}+e^{-{\ga 
k\over{2}}}\right)\widehat{LB^{(+)''}}(k)
+\left(e^{-{3\ga k\over{2}}}+e^{-{\ga 
k\over{2}}}\right)\widehat{L\phi''}(k)
-\widehat{L\mu''}(k) \,, \label{fttpm}\\
-\widehat{{\cal L}t''_-}(k)&=&2\pi\psi_2(k)e^{(\ga-{\pi\over{2}})k}
+\left(e^{{\ga k\over{2}}}+e^{-{\ga k\over{2}}}\right)\widehat{LB^{(-)''}}(k)
+\left[e^{\left({3\ga\over{2}}-\pi\right)k}+e^{\left({\ga\over{2}}-\pi\right)k}
\right]\widehat{L\phi''}(k)+\widehat{{\cal L}\mu''}(k) \,, \non
\ee
which gives
\bea
D_q(k)&=&2\pi\psi_2(k)[e^{-\ga k}-e^{(\ga-{\pi\over{2}})k}]+
\left( e^{{\ga k\over{2}}}+e^{-{\ga k\over{2}}}\right)
[\widehat{LB^{(+)''}}(k)-\widehat{LB^{(-)''}}(k)]\\
&+&2\pi\psi(k)N(e^{i\La k}+e^{-i\La k})\left[e^{-{3\ga k\over{2}}}
+e^{-{\ga k\over{2}}}
-e^{\left({3\ga\over{2}}-\pi\right)k}-e^{\left({\ga\over{2}}-\pi\right)k}\right]
-\oint_{C} \left[ \ln \mu(x) \right]'' e^{ikx} \,. 
\eea

From the definition of $f(x)$ (\ref{deff}), we can derive
\be
\widehat{Lf''}(k)&=&\int_{C_2}dx \Bigg\{ \ln \Bigg[
\sinh(2x-2i\ga) \sinh(2x+2i\ga) 
B^{(+)}(x-{i\ga\over{2}})B^{(-)}(x+{i\ga\over{2}}) \non \\
& & \times \phi(x-{3i\ga\over{2}})\phi(x+{3i\ga\over{2}})\Bigg] 
\Bigg\}'' e^{ikx} \,. \non
\ee
It follows from the identities (\ref{psi}) and (\ref{psi2}) that 
$\widehat{Lf''}(k)$ is given by
\be
\widehat{Lf''}(k)&=&2\pi\psi_2(k)[e^{-\ga k}+e^{(\ga-{\pi\over{2}})k}]+\left[
e^{-{\ga k\over{2}}}\widehat{LB^{(+)''}}(k)+e^{{\ga 
k\over{2}}}\widehat{LB^{(-)''}}(k)
\right]\non \\
&+&2\pi\psi(k)N(e^{i\La k}+e^{-i\La k})\left[e^{-{3\ga k\over{2}}}
+e^{\left({3\ga\over{2}}-\pi\right)k}\right] \,.
\label{ftfresult}
\ee
Similarly, from the definition of $C_{b}(x)$ (\ref{Cb}),
\bea
\widehat{LC''_b}(k)&=&-2\pi\psi_2(k)e^{(\ga-{\pi\over{2}})k}+
e^{{\ga k\over{2}}}\widehat{LB^{(+)''}}(k)-
\left(e^{{\ga k\over{2}}}+e^{-{\ga k\over{2}}}\right)\widehat{LB^{(-)''}}(k)\\
&+&2\pi\psi(k)N(e^{i\La k}+e^{-i\La k})\left[e^{-{\ga k\over{2}}}-
e^{\left({\ga\over{2}}-\pi\right)k}-e^{\left({3\ga\over{2}}-\pi\right)}\right] 
+ e^{-{\ga k\over 2}}\widehat{L\mu''}(k)\,.
\eea
Substituting these results into (\ref{spin1Cdefagain}), we obtain
\bea
C(k)&=&{2\pi N\psi(k)(1-e^{-\pi k})
\over{2\cosh{\ga k\over{2}}}}\left(e^{i\La k}+e^{-i\La k}\right)
+{\left[\widehat{LB^{(+)''}}(k)-\widehat{LB^{(-)''}}(k)\right]
\left(e^{{\pi k\over{2}}}-e^{-{\pi k\over{2}}}\right)
\over{
\left[e^{\left({\pi\over{2}}-\ga\right)k}-e^{\left(\ga-{\pi\over{2}}\right)k}
\right]\left(e^{{\ga k\over{2}}}+e^{-{\ga k\over{2}}}\right)}}\\
&-&{2\pi\psi_2(k)\left(1-e^{-{\pi k\over{2}}}\right)
\left(e^{{\ga k\over{2}}}-e^{-{\ga k\over{2}}}
\right)\over{\left(e^{{\ga k\over{2}}}+e^{-{\ga k\over{2}}}\right)
\left[e^{\left({\pi\over{2}}-\ga\right)k}-e^{\left(\ga-{\pi\over{2}}\right)k}
\right]}}+2\pi k \left(\widehat G(k) +{1\over e^{\ga k\over 
2}+e^{-{\ga k\over 2}}}\right)\,,
\eea
where we have also used the result (\ref{deltaresult}).
From the definition of $B^{(\pm)}(x)$ (\ref{phiBpm}),
\be
\widehat{LB^{(+)''}}(k)&=&
2\pi\psi(k)\left[e^{(\eta_{+}-\pi)k}+e^{(\eta_{-}-\pi)k}\right] \,, 
\non \\
\widehat{LB^{(-)''}}(k)&=&
2\pi\psi(k)\left(e^{-\eta_{+}k}+e^{-\eta_{-}k}\right) \,. \label{Bpmk}
\ee
We conclude that $C(k)$ is given by
\beqa
C(k) &=& 2\pi k \Bigg\{
N\left({e^{i\La k}+e^{-i\La k}\over{2\cosh{\ga k\over{2}}}}\right)
+ {\left[\sinh\left((\eta_{+}-{\pi\over{2}})k\right)+
\sinh\left((\eta_{-}-{\pi\over{2}})k\right)\right]\over{
2\cosh{\ga k\over{2}}\sinh\left(({\pi\over{2}}-\ga)k\right)}}\nonumber\\
&+&{\cosh \frac{\ga k}{4} \sinh\left((3\ga -\pi)\frac{k}{4}\right)\over
\cosh \frac{\ga k}{2}\sinh\left((2\ga -\pi)\frac{k}{4}\right)}
\Bigg\} \,.
\label{Ckfinal}
\eeqa

\subsection{Spin-1: $C_y(k)$}

The quantity $C_y(k)$ is defined by (\ref{Cydef})
\be
C_y(k)={D_T(k)\over{e^{{\ga k\over{2}}}+e^{-{\ga k\over{2}}}}}
+ \widehat{LT''_0}(k) + \widehat{L\mu''}(k)-\widehat{Lf''}(k) \,.
\label{Cydefagain}
\ee
From the definition of $D_T(k)$ (\ref{DTdef}) and the results (\ref{fttpm}),
\bea
D_T(k)&=& e^{-{\ga k\over{2}}}\widehat{Lt''_+}(k)-
e^{{\ga k\over{2}}}\widehat{{\cal L}t''_-}(k)\\
&=&2\pi\psi_2(k)[e^{-{3\ga k\over{2}}}+e^{({3\ga \over{2}}-{\pi\over{2}})k}]
+\left(1+e^{-\ga k}\right)\widehat{LB^{(+)''}}(k)+
\left(1+e^{\ga k}\right)\widehat{LB^{(-)''}}(k)\\
&+&2\pi\psi(k)N(e^{i\La k}+e^{-i\La k})\left[e^{-\ga k}+e^{(\ga-\pi) k}
+e^{-2\ga k}+e^{(2\ga-\pi) k}\right]\\
&+&e^{{\ga k\over{2}}}\widehat{{\cal L}\mu''}(k)-e^{-{\ga 
k\over{2}}}\widehat{L\mu''}(k) \,.
\eea
It follows from (\ref{Cydefagain}) that $C_y(k)$ is given by
\beqa
C_y(k)&=&2\pi\psi_2(k)\left[{e^{-\frac{3\ga k}{2}}
+e^{(\frac{3\ga}{2}-{\pi\over{2}})k}\over{e^{{\ga k\over{2}}}
+e^{-{\ga k\over{2}}}}}\right]+e^{-{\ga k\over{2}}}\widehat{LB^{(+)''}}(k)+
e^{{\ga k\over{2}}}\widehat{LB^{(-)''}}(k)\nonumber\\
&+&2\pi\psi(k)N(e^{i\La k}+e^{-i\La k})\left[e^{({3\ga\over{2}}-\pi)k}
+e^{-{3\ga k\over{2}}}\right] +{e^{{\ga 
k\over{2}}}\widehat{\cal{L}\mu''}(k)-e^{-{\ga k\over{2}}}
\widehat{L\mu''}(k)\over{
e^{{\ga k\over{2}}}+e^{-{\ga k\over{2}}}}} + \widehat{L\mu''}(k)\non \\
&+&2\pi\psi_2(k) -\widehat{Lf''}(k) \non \\
&=& 4\pi k G_2(-k) \,,
\label{Cyfinal}
\eeqa
where, in passing to the last line, we have used the results 
(\ref{G2def}), (\ref{deltaresult}) and (\ref{ftfresult}).

\section{Integration constants}\label{sec:integconst}

We explain here how to determine the integration constants in the 
NLIEs.

\subsection{Spin-$\frac{1}{2}$}\label{sec:spinhalfintegconst}

The main idea is to carefully consider the limit $x\rightarrow 
\infty$. In this limit, the NLIE (\ref{spinhalflatticeNLIE}) becomes
\be
\ln a(\infty) =
G(\infty) \left[ \ln A(\infty) - \ln \bar A(\infty) \right]
+ i\, P_{bdry}(\infty) + i\pi C\,,
\label{spinhalfasympNLIE}
\ee
where $C$ is the constant which is to be determined. The contribution
from the driving term is a multiple of $2\pi i$ (since $N$ is even),
and can therefore be dropped. From the 
definition of $a(x)$ (\ref{defa}) and the fact that $M=N/2$ for the 
ground state, we readily obtain
\be
a(\infty) = e^{i \omega} \,, \qquad \omega \equiv \ga 
\left(H_{+}+H_{-}+2 \right) \,.
\label{defomega}
\ee 
We define the branch of the logarithm such that $| \Im m\ \ln x | \le 
\pi$. It follows that
\be
\ln a(\infty)= i \left(\omega - 2\pi  m \right)\,, 
\label{asympa}
\ee
where $m$ is an integer such that
\be 
-\pi < \omega - 2\pi m < \pi \,.
\label{defm}
\ee
Moreover,
\be
A(\infty)=1+a(\infty)=1+e^{i\omega}=2e^{\frac{i\omega}{2}}\cos \frac{\omega}{2}
=2e^{i(\frac{\omega}{2}+\pi \delta)}\big\vert\cos \frac{\omega}{2}\big\vert \,,
\ee
where $\delta$ is given by
\be
\delta = \left\{ \begin{array}{c@{\quad : \quad} l}
0 & \cos \frac{\omega}{2} > 0 \\
\pm 1 & \cos \frac{\omega}{2}< 0
\end{array} \right. \,.
\label{defdelta}
\ee 
Hence, 
\be
\ln A(\infty) = \ln \left[ 2\big\vert\cos \frac{\omega}{2}\big\vert \right] 
+ i \left( \frac{\omega}{2}+\pi \delta - 2\pi n \right) \,,
\label{asympA}
\ee
and $\ln \bar A(\infty)$ is obtained by complex conjugation. Here
$n$ is another integer such that
\be
-\pi <  \frac{\omega}{2}+\pi \delta - 2\pi n < \pi \,.
\label{defn}
\ee 
We now substitute into (\ref{spinhalfasympNLIE})
the expressions for $\ln a(\infty)$ (\ref{asympa}) and  
$\ln A(\infty)$, $\ln \bar A(\infty)$ (\ref{asympA}), as well 
as the results
\be
G(\infty) = {\pi-2\ga\over 2(\pi-\ga)}\,, \qquad 
P_{bdry}(\infty) = {\pi\over 2(\pi-\ga)}\left(\ga H_{+}+\ga H_{-}+ 
4\ga - 4\pi\right) \,,
\ee 
which follow from (\ref{FTdefG}) and  (\ref{spinhalfPbdry}), respectively.
Solving for $\omega$, we obtain
\be
\omega = \ga H_{+}+\ga H_{-}+ 2\pi \left(C+\delta +2m-2n-2\right) + 
2\ga \left(-C-2\delta -2m+4n+2\right) \,.
\ee
Comparing this result with the definition of $\omega$ in (\ref{defomega}), and 
assuming that $C$ is independent of $\ga$, we obtain a pair of 
equations,
\be
C+\delta +2m-2n-2 = 0 \,, \qquad -C-2\delta -2m+4n+2=1\,,
\ee
which imply
\be
\delta = 2n - 1 
\label{spinhalfdelta}
\ee 
and
\be
 C=3-2m \,. 
\label{spinhalfC}
\ee 

The relations (\ref{spinhalfdelta}) and (\ref{defn}) 
imply that $0 < \omega < 4\pi$. In fact, since $\delta$  
can be only $0$ or $\pm 1$,  (\ref{spinhalfdelta}) implies
$\delta = \pm 1$. It follows from (\ref{defdelta}) that
$\omega$ is further restricted to the interval
\be
\pi < \omega < 3\pi \,.
\label{spinhalfomegarange}
\ee 
Finally, (\ref{defm}) then implies $m=1$, which determines $C$
through (\ref{spinhalfC}), 
\be
C= 1\,.
\ee
Note that the definition of $\omega$ (\ref{defomega}) together with
(\ref{spinhalfomegarange}) imply the domain of boundary parameters
quoted in the text (\ref{bpdomainspin12}).

\subsection{Spin-1}\label{sec:spin1integconst}

In the limit $x\rightarrow \infty$, the spin-1 NLIE becomes
\be
\ln b(\infty) &=& 
G(\infty) \left[ \ln B(\infty) - \ln \bar B(\infty) \right] 
+\frac{1}{2} \ln Y(\infty)
+ i\, P_{bdry}(\infty) + i\pi C\,, \label{spinoneasympNLIE1} \\
\ln y(\infty) &=& 
\frac{1}{2} \left[ \ln B(\infty) + \ln \bar B(\infty) \right] 
+ i\pi C_{y} \label{spinoneasympNLIE2} \,,
\ee
where $C$ and $C_{y}$ are the constants which are to be determined. 
As in the spin-$\frac{1}{2}$ case, we assume that $N$ is even, and 
therefore drop the contribution
from the driving term. From the 
definition of $b(x)$ (\ref{defb}) and $y(x)$ (\ref{yy})
and the fact that $M=N$ for the 
ground state, we obtain
\be
b(\infty) = e^{2i \omega} + e^{4i \omega}\,, \qquad 
y(\infty) = 1+ e^{2i \omega} + e^{-2i \omega}\,, 
\label{asympby}
\ee
where now $\omega$ is defined as 
\be 
\omega \equiv  \eta_{+}+\eta_{-}-\ga  \,.
\label{spin1defomega}
\ee 
We rewrite the expression for $b(\infty)$ as
\be
b(\infty) = 2e^{3i\omega}\cos \omega
=2e^{3i(\omega+\pi \delta_{1})}\big\vert\cos \omega \big\vert \,,
\ee 
where $\delta_{1}$ is given by
\be
\delta_{1} = \left\{ \begin{array}{c@{\quad : \quad} l}
0 & \cos \omega > 0 \\
\pm 1 & \cos \omega< 0
\end{array} \right. \,.
\label{defdelta1}
\ee 
It follows that
\be
\ln b(\infty)= \ln \left[2 \big\vert\cos \omega \big\vert \right]
+ i \left(3\omega +\pi \delta_{1}- 2\pi  m \right)\,, 
\label{asympb}
\ee
where $m$ is an integer such that
\be 
-\pi < 3\omega +\pi \delta_{1} - 2\pi m < \pi \,.
\label{spin1defm}
\ee
Moreover,
\be
B(\infty)=1+b(\infty)=1+e^{2i\omega}+e^{4i\omega}=e^{2i\omega}(1+2\cos 2\omega)
=2e^{i(2\omega+\pi \delta_{2})}\big\vert 1+2\cos 2\omega\big\vert \,,
\ee
where $\delta_{2}$ is given by
\be
\delta_{2} = \left\{ \begin{array}{c@{\quad : \quad} l}
0 & 1+2\cos 2\omega > 0 \\
\pm 1 & 1+2\cos 2\omega < 0
\end{array} \right. \,.
\label{defdelta2}
\ee 
Hence, 
\be
\ln B(\infty) = \ln  \big\vert1+2\cos 2\omega\big\vert 
+ i \left( 2\omega+\pi \delta_{2} - 2\pi n \right) \,,
\label{asympB}
\ee
and $\ln \bar B(\infty)$ is obtained by complex conjugation, where
$n$ is an integer such that
\be
-\pi <  2\omega+\pi \delta_{2} - 2\pi n < \pi \,.
\label{spin1defn}
\ee 
From (\ref{asympby}) we also see that 
\be
y(\infty) = 1+2\cos 2\omega \,,
\label{asympy}
\ee
and therefore
\be
Y(\infty) = 1 + y(\infty) = 2 + e^{2i \omega} + e^{-2i \omega} = 
(e^{i \omega} + e^{-i \omega})^{2} = \big\vert 2\cos 
\omega\big\vert^{2} \,.
\ee 

We now substitute into the first NLIE (\ref{spinoneasympNLIE1})
the above expressions for $\ln b(\infty)$ (\ref{asympb}) and  
$\ln B(\infty)$, $\ln \bar B(\infty)$ (\ref{asympB}), as well 
as the result
\be
G(\infty) = {\pi-3\ga\over 2(\pi-\ga)}
\ee 
which follows from (\ref{Gspin1}),  
and the expression for $P_{bdry}(\infty)$ (\ref{tilderifnty}).
Solving for $\omega$, we obtain
\be
\omega = \eta_{+}+ \eta_{-}+ \pi \left(C-\delta_{1} + \delta_{2} +2m-2n\right) + 
\ga \left(-2C+2\delta_{1} -3\delta_{2} -4m +6n -3\right) \,.
\ee
Comparing this result with the definition of $\omega$ in (\ref{spin1defomega}), and 
assuming that $C$ is independent of $\ga$, we obtain a pair of 
equations,
\be
C-\delta_{1} + \delta_{2} +2m-2n = 0 \,, \qquad 
-2C+2\delta_{1} -3\delta_{2} -4m +6n -3= -1\,.
\label{spinpair}
\ee
These imply that $\delta_{2}=2n-2$, which is an even number. But 
since $\delta_{2}$ can be only $0$ or $\pm 1$, this implies
\be
\delta_{2} = 0 \,, \qquad n=1 \,.
\label{spin1delta2n}
\ee
It follows from (\ref{spinpair}) that 
\be
C= \delta_{1} - 2m +2\,.
\label{spin1C}
\ee 

The relations (\ref{spin1delta2n}) and (\ref{spin1defn}) 
imply that $\frac{\pi}{2} < \omega < \frac{3\pi}{2}$. 
Hence,
\be
\delta_{1} = \pm 1 \,.
\ee 
It follows from (\ref{defdelta2}) and $\delta_{2} = 0$ that
$\omega$ is further restricted to the interval
\be
\frac{2\pi}{3} < \omega < \frac{4\pi}{3} \,.
\label{spin1omegarange}
\ee 
Finally, (\ref{spin1defm}) then implies $\delta_{1}-2m=-3$, which determines $C$
through (\ref{spin1C}), 
\be
C= -1\,.
\ee
Similarly, substituting into the second NLIE (\ref{spinoneasympNLIE2})
the results (\ref{asympB}), (\ref{asympy}), and remembering that 
$\delta_{2} = 0$, we immediately see that
\be
C_{y}=0 \,.
\ee 
Note that the definition of $\omega$ (\ref{spin1defomega}) together with
(\ref{spin1omegarange}) imply the domain of boundary parameters
quoted in the text (\ref{bpdomainspin1}).

\end{appendix}

\end{document}